\documentclass[epj]{svjour}
\usepackage{graphics}
\usepackage{color,amsmath,amssymb,epsfig}

\begin{document}
\title{Transport through  quantum dots: A combined DMRG
and embedded-cluster approximation study}

\author{F. Heidrich-Meisner\inst{1,2}\thanks{\emph{Present address:}  Institut f\"ur Theoretische  Physik C, RWTH Aachen University, 52056 Aachen, Germany; \email{fabian.heidrich-meisner@physik.rwth-aachen.de}} 
\and G.B. Martins\inst{3} \and 
C.A. B\"usser\inst{3,4} \and K.A. Al-Hassanieh\inst{1,2,5} \and 
A.E. Feiguin\inst{6,7} \and
G. Chiappe\inst{8,9} \and E.V. Anda\inst{10} \and E. Dagotto\inst{1,2}
}                     
%
%
\institute{Materials Science and Technology Division, Oak Ridge National Laboratory,
 Tennessee 37831, USA \and
 Department of Physics and Astronomy, University of Tennessee, Knoxville,
 Tennessee 37996, USA
 \and Department of Physics, Oakland University, Rochester, Michigan 48309, USA
 \and
 Department of Physics and Astronomy, Ohio University, Athens, Ohio 45701, USA
 \and
 National High Magnetic Field Laboratory and Department of Physics, Florida State University, Tallahassee, FL 32306, USA
 \and 
 Microsoft Project Q, University of California, Santa Barbara, CA 93106, USA
 \and
 Condensed Matter Theory Center, Department of Physics, University of Maryland, MD 20742, USA
 \and
 Departmento de F\'{\i}sica J.J. Giambiagi, Universidad de Buenos Aires, 
1428 Buenos Aires, Argentina.
\and 
Departamento de
F\'{\i}sica Aplicada,  Universidad de Alicante, San Vicente del Raspeig,
Alicante 03690, Spain
\and
Departamento de  F\'{\i}sica, Pontificia Universidade
Cat\'olica do Rio de Janeiro,
38071 Rio de Janeiro, Brazil
}
\date{Nov. 18, 2008}
%
\abstract{
The numerical analysis of strongly interacting nanostructures
requires powerful techniques.
Recently developed methods, such as the time-dependent density matrix renormalization group (tDMRG) approach  or
the embedded-cluster approximation (ECA), rely on the numerical solution of clusters of finite size.
For the interpretation of numerical results, it is therefore crucial to understand finite-size effects
in detail. In this work, we present a careful finite-size analysis for the examples of one quantum dot,
as well as three serially connected quantum
dots. Depending on ``odd-even'' effects, physically quite different
results may emerge from clusters that do not differ much in their size.  We provide
a solution to a recent controversy over results obtained with ECA for  three quantum dots.
In particular,  using the optimum clusters discussed in this paper, the parameter range in which
ECA can reliably be applied is increased, as we show for the case of three quantum dots.
As a practical procedure, we propose that a comparison of results for static quantities against those 
of quasi-exact methods, such as the ground-state density matrix renormalization group (DMRG) method or exact diagonalization, 
serves to  identify the optimum cluster type.
 In the examples studied here, we find that to observe  signatures of the  Kondo effect in finite systems, the best clusters involving dots and leads
must have a total $z$-component of
the spin equal to zero.
} 
\authorrunning
\titlerunning
\maketitle

\section{Introduction}

The prediction and experimental observation of the Kondo effect in quantum
dots \cite{goldhabergordon98,glazman88,ng88,meir92}
and single-molecule conductors has stimulated considerable interest in strongly correlated
nano-scale systems, as discussed in recent reviews \cite{glazman05,pustilnik05,grobis07}. The related experimental and theoretical efforts
are not only motivated by novel emerging physical phenomena, but
also by the possible technological applications of nanodevices, and in particular
by their transport properties. Experimental results for the conductance through
nanostructures have been reported for single dots, side and linearly coupled two dots,
as well as small molecules \cite{goldhabergordon98,wiel00,park02}.

Stimulated by the rich physics harbored by interacting nanostructures, the field has seen a
rapid development of powerful numerical techniques to obtain the conductance. These include the well-established numerical-renormalization group
approach (NRG) \cite{wilson75,krishnamurthy80a,anders05,bulla08}, various density-matrix renormalization group (DMRG) based methods \cite{bohr06,bohr07,weichselbaum08}, including in particular,  
the recently developed time-dependent DMRG (tDMRG)
\cite{alhassanieh06,schneider06,kirino08,feiguin08c,boulat08,dasilva08}, Quantum Monte Carlo simulations (see, e.g., Ref.~\cite{weiss08}), 
flow-equation approaches \cite{kehrein05}, as well as exact diagonalization combined with an embedding
procedure, the embedded-cluster approximation (ECA) \cite{ferrari99,busser00,chiappe03,busser04}.
We wish to distinguish the latter method from other cluster-embedding approaches used in this
field, sometimes dubbed the embedding method \cite{molina03,rejec03,meden03}.

Beyond numerical methods, there are many analytical techniques to the problem of transport in quantum dots,
ranging from perturbative schemes (see, e.g., Refs.~\cite{kaminski00,rosch01,doyon06}), 
real-time renormalization methods \cite{schoeller00}, and functional renormalization group 
approaches \cite{karrasch06,jakobs07}, to slave-boson calculations \cite{kotliar86}, and variational methods \cite{schoenhammer85,zitko06},
just to mention a few.    Furthermore, density functional theory (DFT)  is widely used to model molecular conductors   (see, {\it e.g.}, Ref.~\cite{marques06} for a review). 
Such DFT-based approaches successfully account for the coupling of the leads to extended molecules by tackling the molecule plus
a substantial part of the leads (see, {\it e.g.}, Refs.~\cite{koentopp06,arnold07}). In that sense, there is a similarity with the embedded-cluster approximation, as will become more obvious in the 
discussion of ECA in Sec~\ref{sec:eca}. While DFT based methods have been shown to correctly describe Coulomb-blockade physics \cite{koentopp06}, interacting resonant level models \cite{schmitteckert08}, and
finite-bias phenomena \cite{arnold07,brandbyge02},  the effects of strong correlations as manifest in  the Kondo effect still seem to be out of
reach for current DFT schemes. In that sense, DFT based methods are complimentary to other approaches mentioned here that emphasize the effect of strong correlations.

Time-dependent DMRG and ECA are particularly pro\-mising
methods to address complex nanoscale systems such as molecular
conductors. Both methods can in principle include the modeling of phonons, and we mention, as an example,
the ECA study Ref.~\cite{alhassanieh05}   on a molecular conductor with a center-of-mass motion. 
Other applications of ECA on timely problems  of  experimental significance include studies of nonlocal spin control in quantum dots
\cite{martins06}, the SU(4) Kondo effect \cite{busser07}, and multilevel quantum dots \cite{martins05}.
Furthermore, it
is one of the advantages of DMRG \cite{white92b,white93,schollwoeck05,hallberg06} and ECA that spatial correlations
 can be analyzed not only within the interacting region, but also throughout the entire system.
Therefore, it is possible to
study how the screening
of the magnetic moment of quantum dots occurs, which relates to the notion of the
so-called {\it Kondo
cloud} \cite{gubernatis87,thimm99,sorensen05,costamagna06}. The computation of extended spatial correlations within
NRG has become possible only recently \cite{borda06}. Moreover, existing codes for both ECA and tDMRG can easily be adapted to different problems, 
and ECA calculations can be performed at low computational costs, {\it i.e.}, the requirements in CPU time needed are rather modest compared to DMRG and NRG.
This is, compared to NRG, of particular advantage in the case of multi-channel problems, such as realized in the case of three quantum dots.

\begin{figure}[t]
\centerline{\input{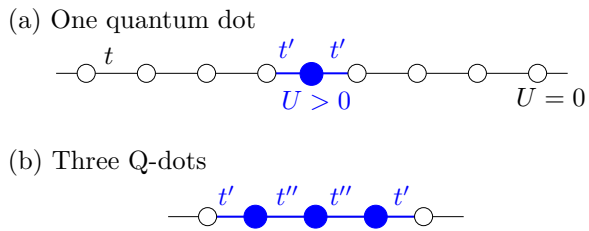}}
\caption{Sketch of the models studied in this work: (a) a single quantum dot
embedded into noninteracting leads; (b) three serially coupled dots embedded into leads.
Full symbols denote dots (namely, sites with a finite onsite Coulomb repulsion $U>0$), while open symbols
represent tight-binding sites. 
}\label{fig:geometry}
\end{figure}

As both DMRG and ECA
are based on real-space schemes,  for the interpretation of
numerical results, it is crucial to understand how subtle many-body effects,
such as the Kondo effect, manifest themselves on finite systems. 
It is the purpose of this work to present a detailed analysis of the properties of
strongly interacting nanostructures and
their finite-size scaling, in both  DMRG and ECA calculations. Our work  first provides a  discussion
of the one-dot   case
 and then  covers three serially coupled quantum dots as well.
 The models are 
depicted in Fig.~\ref{fig:geometry}.

We  discuss how the emergence of Kondo physics 
is reflected  in static properties, such as  spin and charge
fluctuations and spin-spin correlations. In addition, we
discuss the conductance of these structures, obtained with the ECA method,  as a function of model
parameters and gate voltage, which controls the dots' filling. As our
main result, we argue that it is important to
consider the geometric properties and global quantum numbers of
finite systems in order to correctly interpret the results from
exact diagonalization (ED)-based approaches such as ECA. 
The main conclusions are relevant as well for real-time simulations  within
tDMRG performed with  tight-binding leads. We mainly gauge our results against the most established methods, 
the Bethe ansatz solution of the single impurity
problem \cite{andrei80,gerland00}, sum rules valid for Fermi liquids \cite{hewson}, and, for multi-dot structures, NRG \cite{bulla08}.

The existence of odd-even effects in NRG calculations, which use logarithmically
discretized leads (sometimes\\ called Wilson chains, see, e.g.,
Refs.~\cite{weichselbaum08,dasilva08}), has been emphasized by Wilson \cite{wilson75}, with the
important observation that the renormalization procedure of NRG on Wilson chains 
results in two different fixed points, depending on whether the number of NRG iterations is
odd or even. In this work, we discuss odd-even effects that emerge in the finite-size scaling 
of, e.g., the conductance, when the dot is coupled to leads in a real-space representation
\cite{sorensen96}.
 Such even-odd effects and, more generally, the relevance of boundary conditions 
for transport, have been studied before, e.g., for
small chains of dots coupled to wires \cite{kim02}, or quantum box problems \cite{hand06}.
The purpose of our work is to discuss these effects with a perspective on the two methods of 
interest here, i.e., tDMRG and ECA, and most importantly, on the calculation of
conductances for the single-impurity Anderson impurity model and a three-quantum dot model.
We mention that even-odd effects and their influence on the conductance have previously been
studied by static DMRG in Ref.~\cite{molina03}. The model studied there
was spinless fermions  interacting in a constrained region embedded 
in noninteracting leads in a ring geometry.

The case of three
dots \cite{chiappe03,busser04,karrasch06,zitko06,oguri99,oguri05,zitko07,nisikawa06,lobos06,kuzmenko03}
 has attracted considerable attention from
the theoretical side. Most studies find perfect conductance through
the dots at half filling, independently of model parameters such as
hopping matrix elements between the dots and onsite Coulomb
repulsion \cite{karrasch06,zitko06,oguri99,oguri05,zitko07,nisikawa06,lobos06}.
 This result is at odds with
earlier results obtained from ECA since a conductance dip has been
reported in Ref.~\cite{busser04}. Here, we resolve this issue in favor of the picture promoted in Refs.~\cite{karrasch06,zitko06,oguri99,oguri05,zitko07,nisikawa06,lobos06}
and show that the controversial
dip is not due to a general incapability of ECA to capture the essential
physics, but due to quite subtle finite-size effects which have not
been appreciated in some earlier ECA calculations  \cite{busser04}.

The ECA method incorporates  the numerical solution of a small cluster that contains the interacting region -- e.g., a quantum dot --
by means of exact diagonalization.
One example of a relevant global quantum number of such small clusters is  $S_{\mathrm{total}}^z$, the $z$-component of the total spin.
We illustrate that in certain cases, clusters with an overall $S_{\mathrm{total}}^z \not= 0$, at half filling,  and with open boundary conditions
may exhibit qualitatively different properties from those with $S_{\mathrm{total}}^z = 0$.
Furthermore, as a practical procedure, we propose that a quantitative comparison
of static properties, such as charge variations with gate potential, between ECA and ground-state DMRG or the Lanczos algorithm, 
allows us to determine the optimum cluster-type for a given model. A Lanczos solver is part of standard ECA codes as we shall explain later.

The Hamiltonian common to the systems analyzed here consists of three parts:
the noninteracting leads $H_{\mathrm{leads}}$, the coupling between the interacting
region and the leads,  $H_{\mathrm{hy}}$, and the interacting region described by $H_{\mathrm{int}}$:
\begin{eqnarray}
H&=& H_{\mathrm{leads}}+ H_{\mathrm{hy}} + H_{\mathrm{int}}\,,
\label{eq:ham}
\end{eqnarray}
where the leads are described by:
\begin{eqnarray}
H_{\mathrm{leads}} &=& -t \sum_{l=1}^{N_L-1}\lbrack c_{l,\sigma}^{\dagger}  c_{l+1,\sigma} + \mbox{h.c.}\rbrack\nonumber \\
          &&- t\sum_{l=N_L+N_{\mathrm{int}}+1}^{N-1} \lbrack c_{l,\sigma}^{\dagger}  c_{l+1,\sigma}  + \mbox{h.c.}\rbrack \,.
\end{eqnarray}
As usual,  $c_{l,\sigma}^{(\dagger)}$ denotes a fermion annihilation
(creation) operator acting on site $l$, with a spin index
$\sigma=~\uparrow,\downarrow$. Summation over a repeated index $\sigma$ is implied throughout the paper.
$n_{l,\sigma}=c_{l,\sigma}^{\dagger}c_{l,\sigma}$ is the local
particle operator with spin $\sigma$ and $n_l=n_{l,\uparrow}
+n_{l,\downarrow}$ is the  particle number (or charge) operator
on site $l$.  
 The total system size is $N=N_L+N_{\mathrm{int}}+N_R$, where
$N_{L(R)}$ is the number of sites in the left(right) lead and
$N_{\mathrm{int}}$ is the number of interacting sites in the center
of the system.   The
hybridization term can be written as:
\begin{equation}
H_{\mathrm{hy}} = -t^{\prime} \sum_{\sigma}(c_{N_L,\sigma}^{\dagger}  c_{N_L+1,\sigma}
                      +  c_{x,\sigma}^{\dagger}  c_{x+1,\sigma}
        + \mbox{h.c.}) \,,\label{eq:hy}
	\end{equation}
	with $x=N_L+N_{\mathrm{int}}$.
Unless  otherwise stated, $t=1$ is the unit of energy. The quantum of conductance 
in this notation is thus $G_0=2$, where the factor of two is due to the two spin channels.

The rest of the paper is organized as follows. In Sec.~\ref{method}, we introduce the two methods used here, DMRG and ECA. 
In Sec.~\ref{one}, we revisit the case of a single quantum dot. 
We point out that tDMRG calculations result in quite different behaviors, depending on the 
type of open clusters. We qualitatively discuss spin-spin correlations  and calculate the conductance as a function of gate voltage with ECA.
In Sec.~\ref{three}, we present our numerical results
for static properties of three quantum dots obtained from DMRG and ECA calculations, with 
a special focus on their finite-size dependence.
This allows us to identify the optimum clusters for the calculation of dynamic properties and the conductance.
 Section~\ref{conc} provides a summary and conclusions.


\section{Methods}
\label{method}

\subsection{Time-dependent DMRG}
\label{sec:tdmrg0}
For ground-state DMRG calculations, we use the finite-system size algorithm \cite{schollwoeck05} and we give the number of states $M$
used in either the figures' captions or in the text.
The set-up for our adaptive time-dependent DMRG \cite{white04,daley04} calculations of transport properties
has been detailed in Ref.~\cite{alhassanieh06}.  
Here we just repeat that the
current $J(\tau)$ [where $\tau$ denotes time in units of $1/t$] is measured as the current across the link between the leads and 
the interacting region [see Eq.~(\ref{eq:hy})]:
\begin{eqnarray}
j_{x,x+1}  &=&  -i t'\sum_{\sigma}( c^{\dagger}_{x+1,\sigma} c_{x,\sigma}-\mbox{h.c.}) \nonumber \\
	  J&=&(\langle j_{N_L+N_{\mathrm{int}},N_L+N_{\mathrm{int}+1}}\rangle_{\tau}+ \langle j_{N_L,N_L+1}\rangle_{\tau} )/2
\end{eqnarray}
with $x=N_L+N_{\mathrm{int}}$ or $x=N_L$. $\langle \,\cdot \,\rangle_{\tau} $ denotes the expectation value taken in the time-dependent state.
The external bias -- an onsite potential $\pm \Delta V n_i$  applied to the leads -- is $\Delta V\sim 10^{-3}t$.
We use a  Trotter-Suzuki break-up of the time evolution operator in our adaptive tDMRG scheme with a
time step of $\delta \tau\sim 0.05$ \cite{white04,daley04}.
The discarded weight during the time evolution is kept below $10^{-10}$. For the tDMRG data to be discussed in Sec.~\ref{sec:tdmrg}, $M\lesssim 800$ states needed to be
kept to meet this requirement. 
With this choice of tDMRG parameters, the real-time data obtained with a discarded weight of $10^{-10}$ and $5\cdot 10^{-10}$ turn out to be indistinguishable from each other
on the scale of the plots.
Open boundary conditions are imposed in all DMRG calculations. 
As for the notation adopted throughout the manuscript, for our DMRG results, 
$\langle \,\cdot \, \rangle $ denotes the expectation value taken in the ground-state calculated with the method 
in a subspace given by a fixed total number of electrons and a fixed value of $S_{\mathrm{total}}^z$. Further note that on chains with an odd number of
sites and at half filling,  the results on static properties and real-time currents to be presented in the following sections
do not depend on whether we work at $S_{\mathrm{total}}^z=1/2$ or $S_{\mathrm{total}}^z=-1/2$, with the trivial exception  of $\langle S_{i}^z\rangle$, which changes its sign.
For $N$ odd, we hence work in the $S_{\mathrm{total}}^z=1/2$ subspace only.

\subsection{ECA}
\label{sec:eca}

The ECA method  relies on the numerical determination of the ground-state of a cluster with open
boundary conditions. In the following, we briefly sketch details of the method. 

The ECA method tackles  the impurity problem in three steps.
First, the infinite system is artificially cut into two parts: one part {\bf C} that contains
 the interacting region plus as many noninteracting sites of the leads as possible 
(this part will, from now on, be referred to as `the cluster'), and a second
part {\bf R} (the `rest'), consisting of semi-infinite chains positioned at the left and the right in 
relation to the cluster {\bf C}. The number of sites in {\bf C} is denoted by $N_{\mathrm{ED}}$.
Second, Green's functions (GF) for both parts are computed independently: current implementations 
of ECA utilize the Lanczos method to calculate the interacting region's GF, 
while those of the part {\bf R}, being noninteracting, can be computed exactly as well.
In a final step, the artificially disconnected parts are reconnected by means of a 
Dyson equation, which dresses the interacting region's GF. 
This step, the actual embedding, is crucial for capturing the many-body physics associated with 
the Kondo effect. Moreover, although the clusters that can be solved 
exactly by means of a Lanczos routine are rather small, 
being of the order of $N_{\mathrm{ED}}\approx 12$ sites only, 
the embedding step largely compensates for that by dressing the cluster GF and 
effectively extending the many-body correlations, induced by the 
presence of the impurity, into the semi-infinite chains {\bf R}.
 
We now further detail these steps.
 The Hamiltonians of the left and right semi-infinite, tight-binding chains, i.e., the noninteracting {\bf R} part, are described by 
\begin{eqnarray}
H_{\rm sc-L}&=& -t\sum_{l=0,\sigma}^{-\infty} (c_{l\sigma}^\dagger c_{l-1\sigma} + \mbox{h.c.})\nonumber\\
H_{\rm sc-R}&=& -t\sum_{l=N_{\mathrm{ED}}+1,\sigma}^{\infty}\-\-\- (c_{l\sigma}^\dagger c_{l+1\sigma} + \mbox{h.c.}),
\label{sc}
\end{eqnarray}
where in this notation, the sites  labeled by $i=1,\dots, N_{\mathrm{ED}}$ are inside the cluster {\bf C}.
The semi-infinite chains are connected to the  cluster by the 
following term:
\begin{eqnarray}
H_{\rm rest} &=& -V [c_{1\sigma}^\dagger c_{0\sigma} + c_{N_{\mathrm{ED}}\sigma}^\dagger c_{N_{\mathrm{ED}}+1\sigma}] + \mbox{h.c.}\,,
\label{con}
\end{eqnarray}
where $V=t$ is the hopping in the broken link, connecting parts {\bf R} and {\bf C}, 
used for the embedding procedure.  The GF for 
the  cluster {\bf C} and for the semi-infinite chains are calculated at zero temperature. 
Fixing the number of particles $m$ and the $z$-axis projection of the total spin, $S_{\mathrm{total}}^z$, the ground
state and the one-body propagators between all the clusters' sites are calculated.  
For example, $g_{ij}^{(m,S^z_{\mathrm{total}})}$, the undressed Green function for the cluster, propagates a particle 
between sites $i$ and $j$ inside the cluster. 
For the noninteracting, semi-infinite chains, 
the GFs $g_{0}^L$ and $g_{N_{\mathrm{ED}}+1}^R$ at the sites $0$ and $N_{\mathrm{ED}}+1$, located at the extreme ends 
of the semi-infinite chains, at left and right to the cluster, can be easily calculated as well.

The Dyson equation to calculate the dressed GF matrix elements $G^{(m,S_{\mathrm{total}}^z)}_{i,j}$ 
can therefore be written as 
\begin{eqnarray}
G^{(m,S_{\mathrm{total}}^z)}_{i,j} &=& g^{(m,S_{\mathrm{total}}^z)}_{i,j} +
g^{(m,S_{\mathrm{total}}^z)}_{i,1}~V~G^{(m,S_{\mathrm{total}}^z)}_{0,j} \nonumber  \\
   && +
g^{(m,S_{\mathrm{total}}^z)}_{i,N_{\mathrm{ED}}}~V~G^{(m,S_{\mathrm{total}}^z)}_{N_{\mathrm{ED}}+1,j}
\label{Dyson1}
\end{eqnarray}
with
\begin{subequations}
\label{Dyson2}
\begin{eqnarray}
G^{(m,S_{\mathrm{total}}^z)}_{0,j} &=& g_0^L~V~G^{(m,S_{\mathrm{total}}^z)}_{1,j} \\
G^{(m,S_{\mathrm{total}}^z)}_{N_{\mathrm{ED}+1},j} &=& g_{N_{\mathrm{ED}}+1}^R~V~G^{(m,S_{\mathrm{total}}^z)}_{N_{\mathrm{ED}},j}~,
\end{eqnarray}
\end{subequations}
where $V$, as mentioned above, is defined according to $H_{\rm rest}$, Eq.~\eqref{con}. 
Equations~(\ref{Dyson1}) and (\ref{Dyson2}) correspond to a chain approximation, where a locator-propagator
diagrammatic expansion is used \cite{anda81,metzner91}.  Note that ECA is exact in the case of $U=0$.

As mentioned before, the calculation of the propagator $g^{(m,S_{\mathrm{total}}^z)}_{i,j}$ 
requires that fixed quantum numbers $m$ and $S_{\mathrm{total}}^z$ be used. However, after
the embedding procedure, these quantum numbers are not good quantum numbers for the cluster anymore. 
Therefore, we have to incorporate 
processes into the ECA method  that allow for  charge fluctuations in the cluster {\bf C}.
To accommodate this requirement, different implementations of ECA have been devised, either by including  
different spin-mixing strategies \cite{ferrari99,martins06,davidovich02,anda02} 
or by moving the Fermi energy of the leads \cite{chiappe05,aguiar-hualde07}.

For this work, we will  adopt a  spin-mixing strategy different from  
the one used in a previous work analyzing the three-dots geometry \cite{busser04}.
We will  show that this different strategy is instrumental in decreasing 
 finite-size effects observed in Ref.~\cite{busser04}.
The spin mixing proceeds as follows. First, a cluster GF with mixed charge is defined through 
\begin{equation}
g^{(m+p,pS_{\mathrm{total}}^z)}_{i,j} = (1-p)~g^{(m,0)}_{i,j} ~+~ p~g^{(m+1,S_{\mathrm{total}}^z)}_{i,j},
\label{p-equat}
\end{equation}
where $p$ takes values between 0 and 1, and we  assume that $m$ is even, in which case, 
the corresponding $S_{\mathrm{total}}^z=0$. In addition, note that 
for the cluster with charge $m+1$, $S_{\mathrm{total}}^z$ takes values $\pm 1/2$. 
The matrix element $g^{(m+p,pS_{\mathrm{total}}^z)}_{i,j}$ corresponds to a situation where 
the charge in the cluster is between
$m$ and $m+1$. The total charge in the cluster, before embedding, can be easily calculated as
\begin{equation}
q^{pS_{\mathrm{total}}^z}(p) = (1-p)m ~+~ p(m+1) = m+p.
\end{equation}
Using Eqs.~(\ref{Dyson1}) and (\ref{Dyson2}), the dressed
GF $\hat{G}^{(m+p,pS_{\mathrm{total}}^z)}$ is obtained, and from this result, 
the charge in the cluster can be calculated:
\begin{equation}
Q^{pS_{\mathrm{total}}^z}(p) = \frac{-1}{\pi} \int_{-\infty}^{E_F} 
\mbox{Im}~\{\sum_{i}G^{(m+p,pS_{\mathrm{total}}^z)}_{i,i}(\omega)\} d\omega\,,
\end{equation}
where $E_F$ is the Fermi level. The value of $p$ is calculated self-consistently, satisfying
\begin{equation}
Q^{pS_{\mathrm{total}}^z}(p) = q^{pS_{\mathrm{total}}^z}(p).
\label{selfcons}
\end{equation}
If there is spin reversal symmetry, e.g., no magnetic field is applied, one can calculate the total GF as
\begin{equation}
G_{i,j}(p) = {1 \over 2} \sum_{S_{\mathrm{total}}^z=\pm 1/2} G^{(m+p,pS_{\mathrm{total}}^z)}_{i,j}, 
\end{equation}
where $p$ satisfies Eq.~(\ref{selfcons}).
It is important to emphasize that the charge fluctuations taken into account by Eq.~(\ref{p-equat})
are the ones between the cluster and the rest of the system, and not just the ones at the interacting
region described by $H_{\rm int}$. The latter ones involve a very localized neighborhood of the dot and as a consequence,
 are typically already well described on isolated clusters only.

As will be shown in Sec.~\ref{sec:3D_eca}, this alternative way of obtaining the dressed GF leads to improved results in the case of three dots.
Note that in the case of one dot, no substantial difference in the final results obtained from either
the old spin-mixing strategy \cite{ferrari99,martins06,davidovich02,anda02} or the one described here is found. Yet, for the sake of consistency,
we apply the same way of spin mixing as described here throughout the paper.

It is noteworthy to point out that the self-con\-sistent solution for the charge-mixing parameter $p$ 
is either 0 or 1 in the Kondo regime. Therefore, in the Kondo regime, where charge fluctuations are suppressed,  no charge mixing is done at all. 
The parameter $p$ takes a finite value $0<p<1$ as the gate potential drives the system into the mixed-valence regime, and  the purpose of the 
charge mixing is thus mainly to smoothen out the transition from an $m$ electron to an $m\pm 1$ electron ground state, that for the bare
cluster would be a discontinuous one. From a conceptual point of view, the introduction of the parameter $p$ establishes 
a necessary consistency in the theory guaranteeing that the number of electrons in the cluster is the same before and after the embedding process.

For completeness, we give the actual expression that we compute the conductance $G$ from \cite{ferrari99,busser00,busser04b}:
\begin{equation}
G= G_0\,\lbrack t^2\,  \rho_{\mathrm{leads}}(E_F)\rbrack^2 |G_{{l,r}} (E_F)|^2.\,
\label{eq:eca_g}
\end{equation}
Here, $\rho_{\mathrm{leads}}(E_F)$ is the  density of states of the leads at the Fermi-level $E_F$ 
and $G_{l,r} (\omega)$ is the dressed Green function that propagates an electron from the first site $l$ left of the 
cluster to the first site right $r$ of the cluster.

\section{Transport through a single quantum dot revisited}
\label{one}

In this section, we  focus on the application of tDMRG and ECA to
the study of the conductance of a single quantum dot. The interacting portion
of the Hamiltonian (\ref{eq:ham}) is:
\begin{equation}
H_{\mathrm{int}}= U n_{\mathrm{dot},\uparrow}n_{\mathrm{dot},\downarrow} + V_g n_{\mathrm{dot}}\,,\label{eq:1D}
\end{equation}
where the Coulomb repulsion $U$ represents the charging energy, $V_g$ is the gate potential, and
$n_{\mathrm{dot}}=n_{\mathrm{dot},\uparrow}+n_{\mathrm{dot},\downarrow}$ is the dot's charge.
The model is particle-hole  symmetric at $V_g=-U/2$.

As has been shown in Ref.~\cite{alhassanieh06}, tDMRG is
capable of producing zero-temperature transport properties of the single-impurity Anderson model for $U/(2t'^2)\lesssim 4$, such as
the conductance as a function of gate potential   or its dependence
with a magnetic field. In that previous work, it was  noticed that  the
behavior of real-time currents on finite systems with open boundary conditions is
quite different depending on odd-even effects. 
When the number of sites  in each lead is even or odd, we have a 
$S_{\mathrm{total}}^z=1/2$ ground-state, while with an odd number of
tight-binding sites in one lead and an even number in the other, the  system  has an
overall $S_{\mathrm{total}}^z=0$ ground state. Here we present a detailed scaling analysis of static properties
that explains why the latter type of clusters is preferable, as suggested in Ref.~\cite{alhassanieh06}.

Finally, we discuss finite-size effects in the context of ECA, which are of similar
origin, and we provide a quantitative comparison with exact results obtained through the Friedel
sum rule \cite{hewson} and ground-state DMRG.

\begin{figure}[t]
\centerline{\epsfig{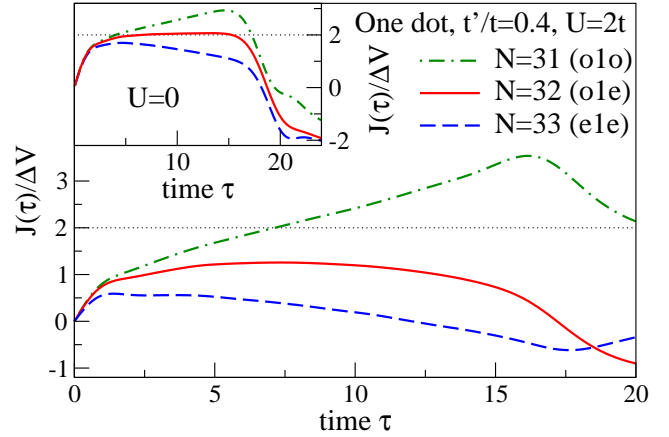}}
\caption{ One dot, $t^{\prime}/t=0.4$, $V_g=-U/2$.
Main panel: current $J(\tau)/\Delta V$ vs. time for $N=31$ ({\it o1o}), $N=32$ ({\it o1e}),
and $N=33$ ({\it e1e}) at $U/t=2$ (tDMRG).
Inset:  current $J(\tau)/\Delta V$ vs. time at $U=0$ for $N=31,32,33$ (exact diagonalization). 
We divide $J(\tau)$ by the bias ($\Delta V=10^{-3}t$). Perfect conductance corresponds to $J(\tau)/\Delta V =2$ (indicated by horizontal, dotted lines).}
\label{fig:one_crr}
\end{figure}

\subsection{Motivation: tDMRG results}
\label{sec:tdmrg}
Let us  motivate our study by discussing results from tDMRG for the current
through a quantum dot at small external biases. 
On an open chain with one embedded dot, we analyze three configurations: (i)
an odd number of sites in both leads ({\it o1o}); (ii) an even number of sites
in both leads ({\it e1e}), (iii) or an overall even number of sites ({\it o1e}).
We repeat some of the main results obtained
in Ref.~\cite{alhassanieh06} for the {\it o1e} cluster-type, 
which  has mostly
been used in that work. First, both at zero and finite $U$, the current is almost constant in time over an extended period in time, allowing us to 
assign a conductance value $G(N)=J(\tau)/\Delta V$ to each chain of finite length. Note that the length in time of such
constant current is limited by  twice the time that it takes the current to be reflected at the open boundaries. Second,
performing a  finite-size scaling of $G(N)$, 
one finds that it  extrapolates, following a simple scaling law of the form $G(N) \sim 1/N$, to the  known exact results from Fermi liquid relations \cite{hewson} 
for $U/2 t'^2\lesssim 4$. 
Third, {\it e1e} clusters have been found to exhibit
a slower convergence and generally lower current values. Our discussion in this section will mostly focus on clusters with 
$S_{\mathrm{total}}^z=1/2$, i.e., {\it e1e} and {\it o1o} ones.

The behavior of a current driven by a small bias is illustrated in Fig.~\ref{fig:one_crr} for clusters
of an intermediate length and at half filling. 
It is important to notice that we choose values for $U$ and $t'$ as well as cluster sizes
such that the finite-size effect we wish to address is clearly visible. 
As a drawback, obviously, a steady state with $G\sim G_0$ is not reached for the three clusters
and system sizes considered in Fig.~\ref{fig:one_crr}. Longer chains are needed for this set of $U$ and $t'$ since
(i) finite-level spacings possibly cut off the Kondo resonance 
and (ii) the time-scale for reaching the steady state \cite{wingreen93,schiller00} may exceed the 
time at which the  current reverses its sign. 

We now turn to this section's main point, the discussion of even-odd effects in the current.
{\it e1e} clusters result in the smallest current, {\it o1e} clusters yield a larger
current and exhibit the aforementioned behavior of being approximately constant in time  ($J(\tau)\approx$const) before the reversal of the current's sign occurs, {\it o1o} clusters
allow for a substantially larger current that even exceeds - if considering the conductance
$G=J/\Delta V$ -  perfect conductance $G_0=2$, but that is never constant in time. These differences are quite striking,
considering that the system sizes used in Fig.~\ref{fig:one_crr} differ by only one site in
the leads. It is worth pointing out that these strong finite-size effects are in principle
not caused by the presence of the Coulomb repulsion on the dot, but can even be seen in the noninteracting case $U=0$,
as displayed in the inset of Fig.~\ref{fig:one_crr}. Note that when the system size is increased,
$J(\tau)/\Delta V$ measured on {\it o1o} clusters approaches perfect conductance from above.
We stress that the strong differences in $J(\tau)$ depending on the cluster type are not 
due to current-ringing \cite{wingreen93,schiller00}, which is most pronounced 
away from half filling ($V_g \neq -U/2$) or at large biases \cite{schneider06,white04,cazalilla02}.

\begin{figure}[t]
\centerline{\epsfig{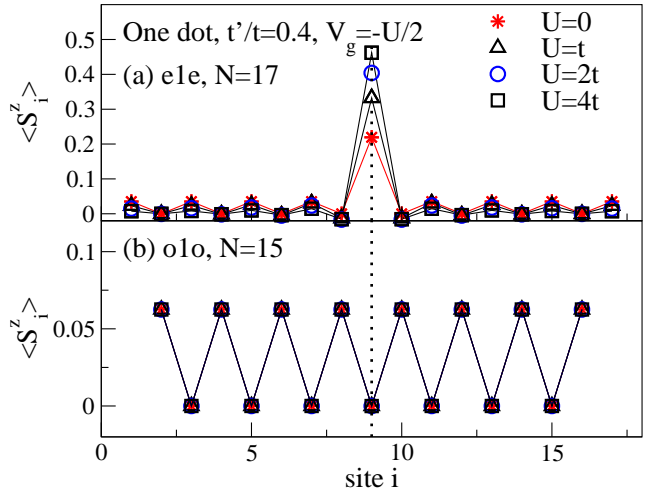}}
\caption{ One dot, $t^{\prime}/t=0.4$, $V_g=-U/2$.
  Spin  density $\langle S_i^{z}\rangle$,
for (a) $N=17$ sites ({\it e1e}) and (b) $N=15$ sites ({\it o1o}) with $U/t=0,1,2,4$.
(a): Even at $U=0$, the $z$-component of the total spin is essentially carried by the central site, {\it i.e.}, the dot.
(b): In the {\it o1o} case, overall, every second site  carries a positive $\langle S_{i}^{z}\rangle$. The dotted vertical lines denote the position of the dot. DMRG data with $M=300$. The results presented here for {\it o1o} and {\it e1e} agree
with Ref.~\cite{sorensen96}. DMRG calculations are all performed in subspaces with a fixed
number of fermions and $S_{\mathrm{total}}^z$, and the symbol $\langle \, \cdot \, \rangle$, here and
throughout, refers to taking the expectation value in the ground state of such subspaces.}
\label{fig:one3d}
\end{figure}

The crucial difference between {\it o1e} on the one hand and the other two clusters
on the other hand is the finite magnetic moment $S^{z}_{\mathrm{total}}=1/2$ present in the system
in the case of an overall
odd number of sites.
As we shall see, this finite moment is not homogeneously distributed over the system.
Let us first explain the suppressed conductance seen in the case of the {\it e1e} clusters.
Figure~\ref{fig:one3d}(a) shows the local spin density $\langle S_i^{z}\rangle$
 for $N=17$ sites and several values of $U/t=0,1,2,4$.
Clearly, for any  $U$, most of the $S_{\mathrm{total}}^{z}$
is carried by the central site, i.e., the quantum dot \cite{sorensen96}.
The reason for this inhomogeneous distribution is easy to understand, considering
the critical tendency of 1D systems towards antiferromagnetic  (AFM) correlations at half filling. For that reason,
the spin cannot be homogeneously distributed over all sites.  We further
observe that on open systems,
 the spins on the first and last tight-binding
site  point up,
which then, with an even number of sites to the left and right of the
dot, favors the spin on the dot to point up as well.
In other words, both leads form an approximate singlet, leaving the remaining moment
$ S_{\mathrm{total}}^z $ on the dot. This is strictly true in the limit of $t'=0$,
where $\langle S_{\mathrm{dot}}^z \rangle=1/2$, while
each lead has a vanishing $\langle S_{\mathrm{total}}^z \rangle$. Increasing $t'$ then reduces  $\langle S_{\mathrm{dot}}^z \rangle$.
Upon increasing
$U$ at a fixed $t'>0$, a larger fraction of  $S_{\mathrm{total}}^{z}$ is pinned at the dot site,
as charge fluctuations that tend
to reduce the local moment are suppressed by increasing $U$.

\begin{figure}[t]
\centerline{\epsfig{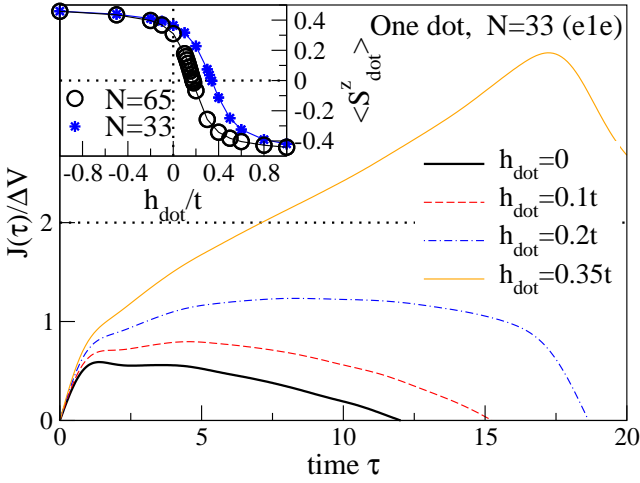}}
\caption{One dot, $t^{\prime}/t=0.4$, $U/t=2$, $V_g=-U/2$.
Main panel: tDMRG data for the current 
$J(\mathrm{\tau})/\Delta V$  vs. time for $N=33$ and several values for a magnetic field $h_{\mathrm{dot}}$
 applied to the dot: $h_{\mathrm{dot}}/t=0,0.1,0.2,0.35$ [see Eq.(\ref{eq:field})].
The inset shows $\langle S_{\mathrm{dot}}^z \rangle $ as a function of 
$h_{\mathrm{dot}}$ for $N=33$ and $65$, calculated with  ground-state DMRG ($M=300$).
}
\label{fig:one1d_field}
\end{figure}

Qualitatively, the situation of an {\it e1e} cluster is equivalent to a quantum dot
in the presence of a magnetic field: locally, the  $Z_2$ symmetry $S_{\mathrm{dot}}^z$ $\leftrightarrow$
$-S_{\mathrm{dot}}^z$ is broken on {\it e1e} clusters.
 A magnetic field is well known to split the Kondo resonance
thus leading to a reduced conductance \cite{hewson}. In order to  support this picture, we
use {\it e1e} clusters and apply a {\it real} magnetic field onto the dot:
\begin{equation}
H_{\mathrm{field}} = h_{\mathrm{dot}} S_{\mathrm{dot}}^z ; \quad S_{\mathrm{dot}}^z =
\frac{1}{2}(n_{\mathrm{dot},\uparrow}-n_{\mathrm{dot},\downarrow}) \,.\label{eq:field}
\end{equation}
As expected, when $\langle S_{\mathrm{dot}}^z\rangle$ decreases upon applying the field $h_{\mathrm{dot}}$ --
as now the field favors a {\it negative} spin in the dot --
the current increases, as we show in Fig.~\ref{fig:one1d_field}. For the parameters of 
Fig.~\ref{fig:one1d_field} ($N=33$, {\it e1e}) we infer
from the figure's inset that the spin projection on the dot vanishes at about
$h_{\mathrm{dot}}\approx 0.34t$.
Consistently, the current measured in tDMRG simulations
at $h_{\mathrm{dot}}\approx 0.35t$ reaches values similar to those observed using {\it o1o}
clusters of comparable system size. 
This  establishes a connection between the dot's polarization and the achievable currents. 
Note further that {\it e1e} clusters behave exactly like {\it o1o} ones if the filling is kept at $(N\pm 1)/N$ (not shown). 

Finally, note that the finite spin projection in the cluster's ground-state  is an issue of parameters. Increasing the coupling to the leads, i.e, $t^{\prime}$,
at fixed $U$ and  system size (say, $N=33$, {\it e1e}),
delocalizes the spin and a larger current is then measured in tDMRG calculations.

For the second type of clusters with a nonzero $S_{\mathrm{total}}^{z}$, {\it o1o},
we find that the spin is equally distributed over every second site \cite{sorensen96}, displayed in
Fig.~\ref{fig:one3d}(b). Since the out-most sites carry a positive
$\langle S_{i}^{z}\rangle>0$, the dot now has a nearly vanishing $\langle
S_{\mathrm{dot}}^{z}\rangle$.
Consistent with the notion of
$\langle S_{\mathrm{dot}}^{z}\rangle\approx 0$, the spin projections on the dot and all other sites are practically independent
of the Coulomb repulsion $U$.

It is important to stress that at fixed odd $N$, both types of
behavior can be produced: when shifting the dot in a $N=33$ cluster
by one site, a {\it 17-dot-15} configuration is obtained. The
current now behaves almost exactly like the one measured on an
$N=31$ ({\it o1o}, {\it 15-dot-15}) cluster and so does the $z$-component
of the spin. The same holds for an asymmetric $N=31$ cluster with a
{\it 16-dot-14} configuration: the dot now carries most of the
$S^z_{\mathrm{total}}$ and the current is suppressed, similar to the
case of $N=33$ ({\it 16-dot-16}, {\it e1e}). The corresponding DMRG
results  are not shown in the figures, but are quantitatively
similar to Figs.~\ref{fig:one_crr} and \ref{fig:one3d}. These
considerations show that indeed the  distribution of
$S_{\mathrm{total}}^z$, which depends on the    geometry  and model parameters, affects the conductance.

We finally  emphasize that the issue of a non-uniformly distributed $\langle S^z_i\rangle$
on half-filled chains with an odd number of sites and constant hopping matrix elements in the leads  
also emerges in other interacting systems that do not feature Kondo physics, such as transport through a Mott insulator
or an even number of quantum dots \cite{alhassanieh-unpub}. Its influence on the finite-size scaling of transport properties 
is most significant in Kondo problems, though. 

 \begin{figure}[t]
\centerline{\epsfig{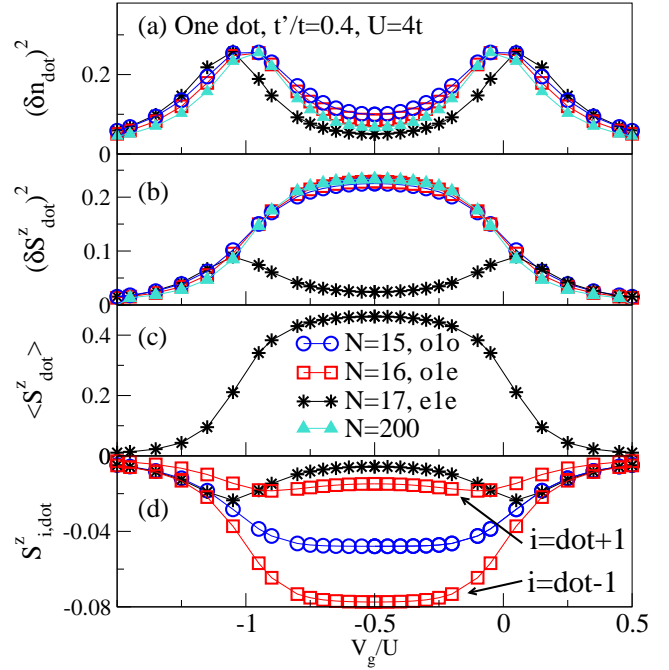}}
\caption{One dot, static properties computed with DMRG as a function of gate potential $V_g/U$ at $t^{\prime}/t=0.4$, $U/t=4$,
comparing different cluster types: {\it o1o} ($N=15$, circles), {\it o1e} ($N=16$, squares),
and {\it e1e} ($N=17$, stars). Triangles in (a) and (b) are for a large system of $N=200$ sites ($M=300$ DMRG states).
 (a)  Charge fluctuations  $(\delta n_{\mathrm{dot}})^2$; (b)
spin fluctuations $(\delta S^z_{\mathrm{dot}})^2$; (c)
local spin density $\langle S^z_{\mathrm{dot}} \rangle$ on the dot; and
 (d)
spin-spin correlations $S_{i,\mathrm{dot}}^z$
between the dot and the first neighboring site in the leads.
For {\it o1e} clusters such as $N=16$, there is an asymmetry between the left and the right
lead: the dot is mostly screened by the left lead, i.e., the one that has an odd number of
sites.
}\label{fig:1D_static}
\end{figure}

\subsection{Static properties: Gate potential dependence}
\label{sec:1d_gate_potential}

To render the line of reasoning outlined in  Sec.~\ref{sec:tdmrg}
more quantitative we proceed with a systematic discussion
of static properties that are crucial to characterize a dot in the Kondo regime:
spin and charge fluctuations on the dot, as well as spin-spin correlations between
the dot and the leads. For this purpose, we focus on the parameter set $t^{\prime}/t=0.4$
and $U/t=4$ and the dependence on the gate potential, keeping the full system at half filling.

First of all, let us recall some of the hallmark
 features of the single-impurity Anderson model at $T=0$ \cite{bulla08,hewson}:
(i) a slowly varying  charge $\langle n_{\mathrm{dot}} \rangle\approx 1$ on the dot
in a broad gate-potential window $-1\lesssim V_g/U \lesssim 0$ for $\Gamma \ll U$,
 where $\Gamma$ denotes the
hybridization parameter.
This property, i.e., the almost constant charge, becomes more
pronounced as $U/\Gamma$ increases. (ii) Enhanced spin fluctuations on the dot.
(iii) Large charge fluctuations at the charge degeneracy points
$V_g/U=-1,0$, but a suppression of charge fluctuations in between.
While the spin fluctuations at the particle-hole symmetric point
increase with increasing $U/\Gamma$, the charge fluctuations
decrease at the same time. Still, the conductance is always perfect,
since the scattering of conduction electrons screens the magnetic
moment, giving rise to the large spin fluctuations. We
characterize charge and spin fluctuations via:
\begin{eqnarray}
(\delta S^z_{\mathrm{dot}})^2 &=& \langle (S_{\mathrm{dot}}^z)^2\rangle- \langle
S_{\mathrm{dot}}^z\rangle^2\,;\\
(\delta n_i)^2 &=& \langle n_i^2\rangle-\langle n_i\rangle^2\,.
\label{eq:spin-fluct}
\end{eqnarray}

Properties (ii) and (iii) of the single-impurity Anderson model can nicely be seen in
Figs.~\ref{fig:1D_static}(a) and (b):
{\it o1o} and {\it o1e} clusters exhibit  a broad maximum in the spin fluctuations around
the particle-hole symmetric point $V_g=-U/2$, while
charge fluctuations are suppressed [Figs.~\ref{fig:1D_static}(a) and (b)]. The maxima in $(\delta n_{\mathrm{dot}})^2$ vs. $V_g/U$
at $V_g/U=-1,0$ are due to the charge degeneracy between states with $n_{\mathrm{dot}}=0$ and 1
and  $n_{\mathrm{dot}}=1$ and 2, respectively. For comparison, DMRG results
for $N=200$ sites (triangles) are included in Figs.~\ref{fig:1D_static}(a) and (b).

As for the {\it e1e} clusters, we emphasize that the main consequence of the  finite spin projection
$\langle S_{\mathrm{dot}}^z \rangle$ present for $-1 \lesssim V_g/U\lesssim 0$ [see  Fig.~\ref{fig:1D_static}(c)]
on the dot is
a reduction of spin fluctuations [see Fig.~\ref{fig:1D_static}(b)].
A Kondo-resonance, however, which is at the heart of perfect transmission through
a quantum dot at the particle-hole symmetric point,
emerges as a consequence of virtual charge fluctuations due to scattering of conduction
electrons off the impurity
that cause substantial spin fluctuations on the dot, thus screening its moment \cite{hewson}.

\begin{figure}[t]
\centerline{\epsfig{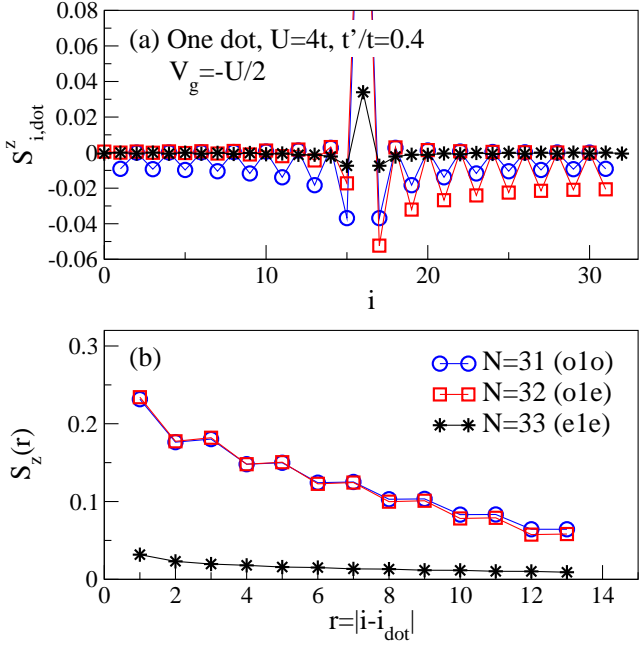}}
\caption{ One dot, $t^{\prime}/t=0.4$, $U/t=4$, $V_g=-U/2$.
(a) Spin-spin correlations vs. site,
(b) integrated spin-spin correlations $S_z(r)$ [see Eq.~(\ref{eq:sr})] vs. distance $r$.
DMRG data are shown for $N=31$ (circles, {\it o1o}), $N=32$ (squares, {\it o1e}),
$N=33$ (stars, {\it e1e}); $M=300$. See the text in Sec.~\ref{sec:1d_gate_potential} for a discussion.
}\label{fig:corr}
\end{figure}

Apart from spin fluctuations, and equally   importantly,
spin-spin correlations between the dot and the leads are affected as well. Figure~\ref{fig:1D_static}(d)
shows the spin-spin correlations between the dot and the first site in the leads:
\begin{equation}
S_{ij}^z = \langle S_i^{z} S_{j}^z \rangle -  \langle S_i^{z}\rangle \langle S_{j}^z \rangle\,,\label{eq:corr}
\end{equation}
which are  substantially smaller around the particle-hole symmetric point $V_g=-U/2$
on {\it e1e} clusters than for other clusters.
These two features illustrate the failure of {\it e1e} clusters to harbor precursors of Kondo physics as long as system sizes are small. We shall discuss the finite-size
scaling  below.

 Let us next discuss how the emergence of screening is reflected in the spatial
 extension of spin-spin correlations. DMRG results for spin-spin correlations [see Eq.~(\ref{eq:corr})],
 measured away from the dot,
 and for their integral $S_z(r)$,
\begin{equation}
S_z(r) = \sum_{i\, \mathrm{with} \atop r'(i)\leq r} S_{i,i_{\mathrm{dot}}}^z; \quad
r'(i)=|i-i_{\mathrm{dot}}|\,,\label{eq:sr}
\end{equation}
are displayed in Fig.~\ref{fig:corr}(a) and (b), respectively,  for
$N=31$ ({\it o1o}), $N=32$ ({\it o1e}), and $N=33$ ({\it e1e}).
We first observe from Fig.~\ref{fig:corr}(a) that, while the leads are symmetrically polarized in the case of
$N=31$, {\it o1e} clusters mainly polarize one lead, namely the one that has an odd
number of tight-binding sites. In this case, an {\it approximate} singlet is formed with this lead.
When integrated over distance, the spin correlations behave quite similarly for these two cluster types, i.e., {\it o1e} and {\it
o1o}.
In contrast to that and consistent with the notion of the absence of Kondo physics on small {\it e1e} clusters,
only small spin-spin correlations spread out into the leads for these clusters.

\begin{figure}[t]
\centerline{\epsfig{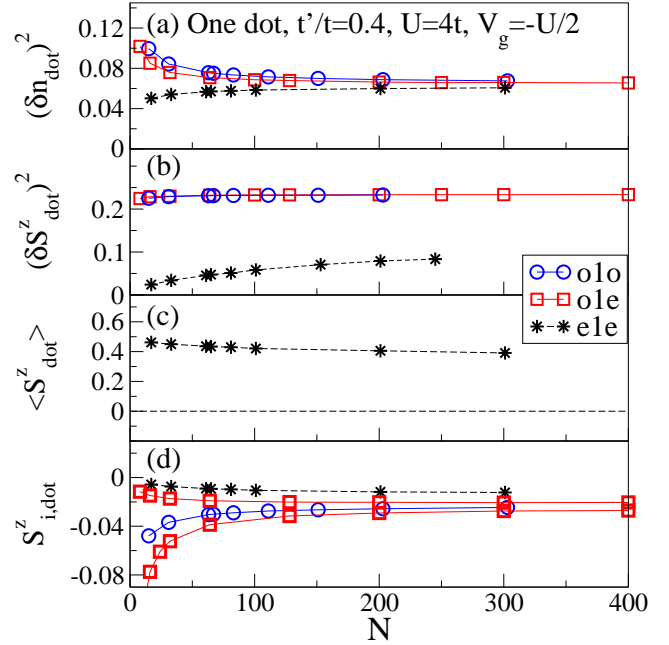}}
\caption{ One dot, $t^{\prime}/t=0.4$, $U/t=4$, $V_g=-U/2$.
Finite-size scaling analysis of static properties: (a) charge fluctuations on the dot; (b)
spin fluctuations on the dot; (c) local spin $\langle S_{\mathrm{dot}}^z \rangle$ on the dot (compare Ref.~\cite{sorensen96});
(d) spin-spin correlations between the dot and its neighboring sites. In (d), the upper {\it o1e} curve is for $i$ in the right lead ($i={\mathrm{dot}}+1$),
while the lower one is for $i$ in the left lead ($i={\mathrm{dot}}-1$).
DMRG results ($M\leq 500$) are shown for {\it o1o} (circles), {\it o1e} (squares), and {\it e1e} clusters (stars).
 }\label{fig:fs}
\end{figure}

\subsection{Static properties: Finite-size scaling}
\label{sec:1d_fs}

We proceed by discussing  the finite-size scaling of
static properties. Our DMRG results for system sizes $15 \leq N\leq 400$ are displayed in
Fig.~\ref{fig:fs}. The main observations are: (i) spin and charge fluctuations converge the
fastest on {\it o1e} clusters; (ii) in the case of {\it e1e}, the dot carries a large, finite
$\langle S_{\mathrm{dot}}^z \rangle$ that decays slowly with $N$, suppressing spin fluctuations and
spin-spin correlations with the dot's neighboring sites; (iii) antiferromagnetic spin-spin correlations
$S_{i,{\mathrm{dot}}}^z$ are enhanced in the case of ${\it o1o}$ clusters on small systems, while
on {\it o1e} clusters, mainly the lead with an odd number of sites develops substantial AFM correlations with the dot.

\subsection{ECA results for the conductance}
\label{sec:1dot_eca}

Early studies \cite{chiappe03} using this technique have phenomenologically proposed that for one dot, the best results
are obtained  when it is neighbored by an odd number of sites ({\it o1o}), while
 {\it e1e} clusters  exhibit a Coulomb-blockade like behavior. Using  insights of
Secs.~\ref{sec:1d_gate_potential} and \ref{sec:1d_fs}, we can now explain this observation on a more rigorous basis.
Small {\it e1e} clusters typically locate the $S_{\mathrm{total}}^z=1/2$ to a large extent
onto the quantum dot, thus giving rise to a split resonance.
Consistently, we find
a charge gap
in the local density of states (LDOS). While the  embedding procedure manages to redistribute
spectral weight in the LDOS quite well \cite{anda08}, it does not close the charge gap given by the
exactly solved interacting region.

\begin{figure}[t]
\centerline{\epsfig{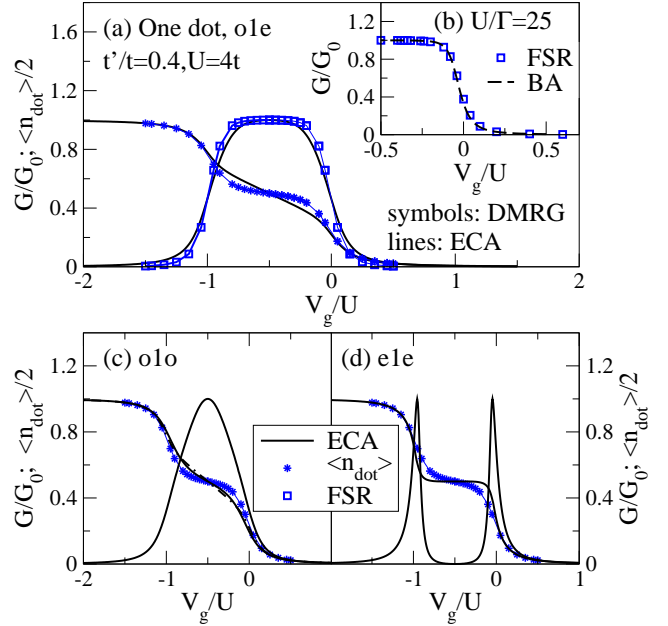}}
\caption{ One dot, $t^{\prime}/t=0.4$,  $U/t=4$.
(a) Conductance  $G$ and charge $\langle n_{\mathrm{dot}}\rangle $ vs. gate potential $V_g/U$
 for {\it o1e} (ECA, $N_{\mathrm{ED}}=12$, solid lines), and DMRG results for the charge and $G$
 obtained through the FSR (circles and squares, respectively).
ECA results  for an {\it 1o} cluster with $N_{\mathrm{ED}}=10$ sites
 are indistinguishable from the {\it o1e} curve.
 (b): Comparison of conductance  obtained from the Friedel sum rule using DMRG
data for  the charge $\langle n_{\mathrm{dot}}\rangle$ ($N=300$, $M=500$, $U/t=0.5$, $t^{\prime}/t=0.1$)
with  Bethe-ansatz results from Ref.~\cite{gerland00}.
(c) Conductance  $G$ and charge $\langle n_{\mathrm{dot}} \rangle$ vs. gate potential $V_g/U$
 for {\it o1o} (ECA, $N_{\mathrm{ED}}=11$, solid lines), and DMRG results for the charge
 (circles).
 (d) Conductance  $G$ and charge $\langle n_{\mathrm{dot}}\rangle$ vs. gate potential $V_g/U$
 for {\it e1e} (ECA, $N_{\mathrm{ED}}=9$, solid lines), and DMRG results for the charge
 (circles).
}\label{fig:1dot_eca}
\end{figure}

For illustration, we show ECA results for one dot at $t^{\prime}/t=0.4$, $U/t=4$ and all three cluster
types in Fig.~\ref{fig:1dot_eca}(a), (c), and (d). The conductance is computed from Eq.~\eqref{eq:eca_g}. The figure also contains the conductance as obtained
from the Friedel sum rule (FSR)  using the charge computed with static DMRG \cite{hewson}:
\begin{equation}
G= G_0 \sin^2(\pi \langle n_{\mathrm{dot}}\rangle/2)\,,
\label{eq:fsr}
\end{equation}
where $G_0=2$ due to spin degeneracy. Here we  claim
that the charge $\langle n_{\mathrm{dot}}\rangle$ can be obtained to high precision
from ground-state DMRG. To support this we show a comparison for the
conductance derived from Eq.~(\ref{eq:fsr}) with exact Bethe-ansatz
results \cite{gerland00} in Fig.~\ref{fig:1dot_eca}(b) for
$U/\Gamma=25$. Note that in the wide-band limit $U< 4t$, the
hybridization parameter is $\Gamma=2 \pi t^{\prime 2}
\rho_{\mathrm{lead}}$, and since we work with semi-infinite-leads,
$\rho_{\mathrm{lead}}=1/(\pi\,t)$ \cite{busser04b}. Our DMRG results
are for $U=0.5t$, $t^{\prime}/t=0.1$. We find excellent agreement
with the results from Ref.~\cite{gerland00} in the limit of a large $U/\Gamma$,
which justifies the use of static DMRG to benchmark the ECA results at smaller 
values of $U/\Gamma\sim 12$.

Clearly, the {\it o1e} clusters are closest to the FSR as shown in the main panel of Fig.~\ref{fig:1dot_eca}(a).
  In Fig.~\ref{fig:1dot_eca}(d), we further observe a Coulomb blockade type-of behavior for {\it e1e} clusters.
Regarding the {\it o1o} cluster, although it qualitatively provides
the correct physical description, the conductance
plateau comes out too narrow as the charge around the particle-hole
symmetric point $V_g=-U/2$ varies too fast as compared to the DMRG
result [see Fig.~\ref{fig:1dot_eca}(c)]. Note that the ECA results
for the charge $\langle n_{\mathrm{dot}} \rangle$ as displayed in
Fig.~\ref{fig:1dot_eca} are obtained {\it after} embedding by
integrating over the imaginary part of the dot's onsite Green's
function.

A further improvement of results can be achieved by  using a
transformation onto bonding and anti-bonding orbitals in the Hamiltonian
for one dot, Eqs.~(\ref{eq:ham})--(\ref{eq:hy}), and (\ref{eq:1D}) (see, e.g., Refs.~\cite{wilson75,krishnamurthy80a} and references therein):
\begin{equation}
b_{i,\sigma}^{\dagger}= \frac{1}{\sqrt{2}}\lbrack c_{i_L,\sigma}^{\dagger}+c_{i_R,\sigma}^{\dagger}\rbrack;
\quad a_{i,\sigma}^{\dagger}= \frac{1}{\sqrt{2}}\lbrack c_{i_L,\sigma}^{\dagger}-c_{i_R,\sigma}^{\dagger}\rbrack\,.
\label{eq:trafo}
\end{equation}
Here, the index $i$ measures the distance away from the dot and
$i_L, i_R$ are the corresponding indices in the two semi-infinite
leads. The transformation decouples the dot from the anti-bonding states, leaving it 
coupled only to the bonding states. This is technically a great simplification as 
leads almost twice as large can now be treated. While such
transformation cannot be used for the spatially asymmetric {\it o1e}
clusters, it works for {\it e1e} and {\it o1o}. We observe that
exactly the same result is found for {\it e1e}, comparing, say, the
full $N=9=${\it 4-dot-4} with the symmetrized {\it dot-4} clusters.
Here, the spin still cannot be screened in a singlet state. For {\it
o1o}, the results substantially improve, as now, effectively a {\it
dot-odd} ({\it 1o}) cluster is solved that has an overall
$S_{\mathrm{total}}^z=0$ ground-state, allowing for singlet
formation. This gives results similar  to those
obtained from an {\it o1e} configuration with less computational effort, as   a smaller
cluster is diagonalized.

Our observations lend strong support to the idea that
when the global quantum numbers of the exactly solved cluster
 allow for a singlet state to be the ground state -- as is the case
for {\it o1e} and {\it 1o} clusters -- then a precursor of the
Kondo effect can be seen even on small ED clusters, with a
favorably fast convergence with system size.

Finally, Figs.~\ref{fig:1dot_eca}(a), (c), and (d) suggest a  way to gauge the quality of ECA data
 even in a situation where no exact results for the conductance are available.
We see that the cluster type that produces the best charge vs gate potential curve
yields  the best $G$ vs. $V_g$ curve as well. Thus, performing a finite-size scaling analysis of properties such as the charge or spin fluctuations
using either a Lanczos routine (which is a part of standard ECA codes, see Sec.~\ref{sec:eca}) or DMRG is well
suited to identify the {\it optimum} cluster type to be used in conductance calculations in ECA.

\subsection{One quantum dot: Summary}

Let us summarize our observations from this Sec.~\ref{one}. We can understand the behavior of
reflection-symmetric
 clusters with an odd number of sites and an   $S_{\mathrm{total}}^z=1/2$ ground state in terms of rigid spins $\langle S_{i}^z\rangle$ pinned to
 certain sites.
In the case of {\it e1e}, the dot is mostly polarized and carries a finite magnetization.
This is qualitatively equivalent to the  application of a magnetic field
that causes Zeeman splitting of the Kondo
resonance and, thus, a suppression of the conductance \cite{hewson}.
For the case of {\it o1o} clusters, the leads are polarized, with only every second site carrying a finite magnetization \cite{sorensen96}. For comparable
system sizes, a substantially larger current is observed on these clusters, which, in tDMRG simulations, however, does not
exhibit a current that is constant in time 
for $t^{\prime} \not= t$.

Our analysis further suggests that either {\it o1e} or {\it
1o} clusters are best suited to study the emergence of Kondo physics
with techniques such as  tDMRG or ECA, while on system sizes
accessible to these techniques, {\it e1e} clusters yield
qualitatively different results with essentially a Coulomb-blockade like 
behavior. The overall conclusion is that in order to provide a
reliable description of the many-body effects at the Fermi level,
clusters with a $S_{\mathrm{total}}^z=0$ ground state are necessary  in calculations with these two methods.
Among these, the advantage of {\it 1o} clusters is that larger system
sizes can be accessed. Finally, we stress that no matter what the cluster type is, if one
was able to go to extremely long chains, the same physics would be recovered 
\cite{sorensen96}.


\begin{figure}[t]
\centerline{\epsfig{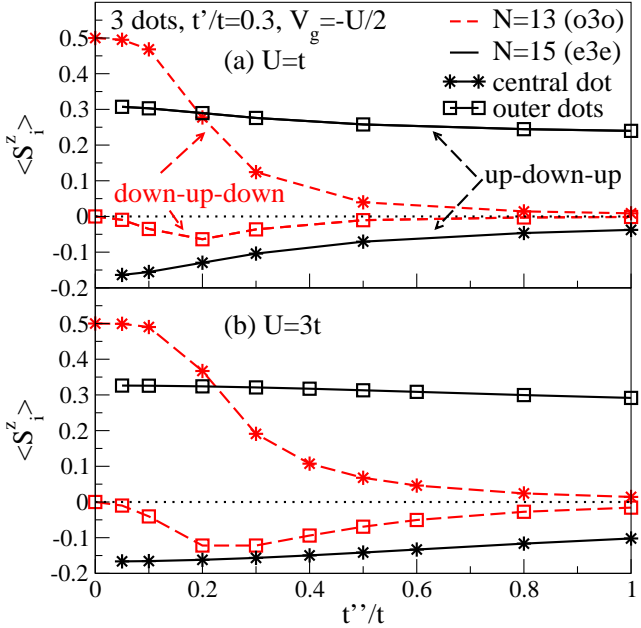}}
\caption{ Three dots, $t^{\prime}/t=0.3$, $U/t=1$ (a) and $U/t=3$ (b).
(a) Spin projection $\langle S_{i}^{z}\rangle$ vs. $t^{\prime \prime}/t$ for the 
outer dots (squares) and the central  dot (stars) for $N=13$ ({\it o3o}, dashed 
lines) and $N=15$ ({\it e3e}, solid lines) clusters. (b) Same as in (a), but for 
$U/t=3$. Note that the magnitude of the fixed spins for both clusters ({\it o3o} 
and {\it e3e}) increase with the value of $U$.
On {\it o3e} clusters, $\langle S^z_{\mathrm{i}}\rangle =0$ (DMRG; $M=300$).}
\label{fig:3D_spin}
\end{figure}

\section{Three Dots}
\label{three}

We now turn to a more complicated case, three serially coupled dots, as shown in Fig.~\ref{fig:geometry}(b).
Apart from the interest in the emergence of Kondo physics, many quantum dots arranged in an array can be
considered an interpolation between a single localized impurity and a bulk Mott insulator. In the first case,
the transmission at half filling is perfect, but zero in the later due to the presence of the Mott gap \cite{lieb68}.

The interacting region is  described by
\begin{eqnarray}
H_{\mathrm{int}} &=&-t^{\prime \prime}\sum_{i=1}^{2}
	(c_{i,\sigma}^{\dagger}c_{i+1,\sigma}+ \mbox{h.c.})\nonumber \\
         &&+ U \sum_{i=1}^{3} n_{i,\uparrow} n_{i,\downarrow} + \sum_{i=1}^{3}V_g n_i\,.
        \label{eq:three_dots}
\end{eqnarray}
To simplify the notation, we will refer to the dots as {\bf D1}, {\bf D2}, and {\bf D3},
where {\bf D2} is the central dot.
We focus on the behavior at the particle-hole symmetric point unless stated otherwise.
In the case of three dots, we distinguish between odd-3-odd ({\it o3o}),
odd-3-even ({\it o3e}), and even-3 even ({\it e3e}) clusters.

Following Refs.~\cite{zitko06,zitko07}, for large $t^{\prime \prime}\gtrsim U/\sqrt{2}$, 
the three dots can be viewed as a molecule with, at half filling, two
electrons occupying its lowest state, and a resulting $S_{\mathrm{total}}^z=1/2$ from the third electron.
For intermediate $t^{\prime \prime}\sim 0.3 U$, 
it has been suggested that the three dots behave 
as a linear antiferromagnet due to strong AFM spin-spin correlations between them.
Finally, at small $t^{\prime \prime}\ll U$, the system enters a very subtle two-stage Kondo regime.
First, and below a Kondo scale $T_K^{(1)}$, the outer dots form a Kondo singlet with their
adjacent leads. Then, at temperatures below a second Kondo scale $T<T_K^{(2)}<T_K^{(1)}$,
the central dot is screened by the quasi-particles of the two Fermi liquids formed
by the outer dots and their leads.
Although $T_K^{(2)}$ is orders of magnitude smaller than $T_K^{(1)}$ \cite{zitko06,cornaglia05}, which renders
an experimental confirmation of the effect a challenging task, nevertheless
the two-stage process conceptually plays an important role for extremely low temperature physics.
 In all three cases and at temperature $T=0$, the system possesses
a spin $S=1/2$ Kondo ground state.

Concerning the issue of the conductance, several studies using a variety of approaches such
as perturbation theory \cite{oguri99}, NRG \cite{zitko07,nisikawa06}, and
functional renormalization group \cite{karrasch06} report perfect conductance at the particle-hole
symmetric point due to the
Kondo effect. An earlier ECA work on this model \cite{busser04}  is at odds with
this conclusion. In particular, an exact zero of the conductance at the particle-hole
symmetric point has been found for all values of $t^{\prime \prime}$ studied, i.e., $t^{\prime \prime}\lesssim 0.5t$.

We show  that  the vanishing of $G$ at $V_g=-U/2$ at $t^{\prime\prime}/t\lesssim 0.
3$ is due to the  choice of {\it o3o} clusters for the  ECA calculations of 
Ref.~\cite{busser04}.
At large $t^{\prime\prime}> t^{\prime}$, this dip is an artefact of certain approximations taken in
the embedding procedure: if we perform the ECA according to the prescription of Sec.~\ref{sec:eca} without
mixing Green functions for spin up and and down before the embedding is carried out, this artifact will disappear on {\it o3o} clusters.

We shall
outline below that the optimum cluster type -- in the sense of fastest convergence of static properties with system size  --
is {\it o3e}. On this cluster type, no spurious dip at $V_g=-U/2$  exists, 
independent of the spin-mixing strategy adopted in ECA calculations, and the ECA results then are in agreement with
the conclusions of Refs.~\cite{karrasch06,oguri99,zitko07,nisikawa06}.

The section is organized as follows: We first analyze static properties and their finite-size scaling in Sec.~\ref{sec:3d_fs}.
Section~\ref{sec:3D_eca} contains our ECA results, obtained  following the procedure discussed in Sec.~\ref{sec:eca},
for the conductance. In Sec.~\ref{sec:o3e}, we briefly discuss the results one obtains in the regime of $t^{\prime\prime}\lesssim t'$ using {\it o3e}  clusters.

\begin{figure}[t]
\centerline{\epsfig{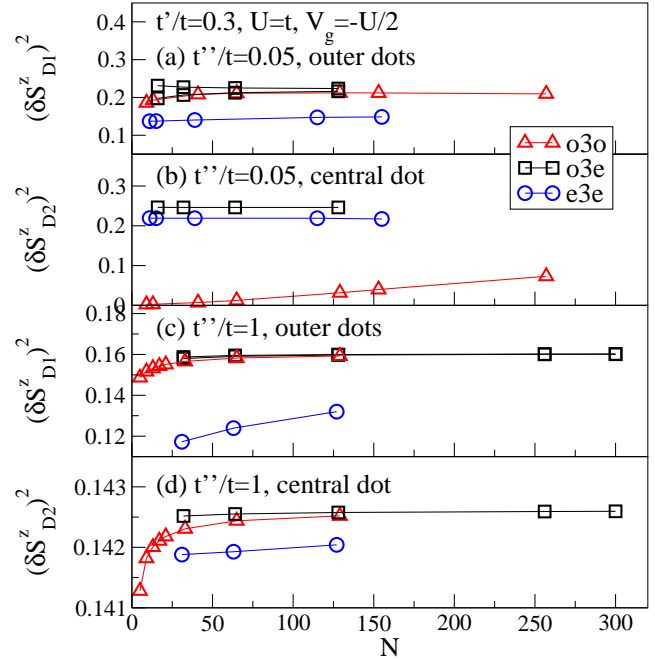}}
\caption{ Three dots, $t^{\prime}/t=0.3$, $U/t=1$.
Finite-size scaling of spin-fluctuations on the dots for $t^{\prime \prime}/t=0.05$ [(a),(b)]
and $t^{\prime \prime}/t=1$ [(c),(d)]. DMRG results  for all three cluster types:
{\it o3e} (squares), {\it e3e} (circles), and {\it o3o} (triangles) [$M\leq 500$].
}\label{fig:3D_scaling}
\end{figure}

\subsection{Static properties: Finite-size scaling}
\label{sec:3d_fs}
Let us first consider  the three dots isolated from the leads ($t^{\prime}=0$) at $U=t$
as a function of $t^{\prime \prime}/t$, which will help to understand {\it e3e} and {\it o3o}
clusters.
In the $S_{\mathrm{total}}^z=1/2$ subspace, the two lowest-lying states
are both doublets ($S=1/2$), however, with different spin projections $\langle S_{i}^z\rangle$ on the dots.
The ground state has -- schematically -- an up-down-up pattern, while the first excited
state realizes an up-up-up structure, with most of the $S_{\mathrm{total}}^z=1/2$ on the
central dot. The actual distribution of $\langle S_{i}^z \rangle$ depends on
$t^{\prime \prime}/U$ for a given $t^{\prime}$,  but when the isolated
cluster is embedded in leads with an overall odd number of sites (i.e., all sites in the leads plus the three  dots), the two low-lying states
are mixed with different weights: on {\it o3o} clusters, an down-up-down configuration
is preferred, while on {\it e3e} ones, the original ground-state pattern (up-down-up)
prevails. 
We illustrate this in Fig.~\ref{fig:3D_spin} where we plot $\langle S_{i}^z\rangle $
vs. $t^{\prime \prime}/t^{\prime}$ for $N=13$ ({\it o3o}) and $N=15$ ({\it e1e}).
We further see that, as a function of increasing $t^{\prime \prime}/t$, the spin projection on the central dot (displayed with stars in the figure)
is substantially reduced and moved to the leads in the case of {\it o3o}.
The large $\langle S_{i}^z \rangle$ located on the outer dots (displayed with squares in the figure)
seen in the case of {\it e3e} decays much slower as a function of $t^{\prime \prime}/t$.

\begin{figure}[t]
\centerline{\epsfig{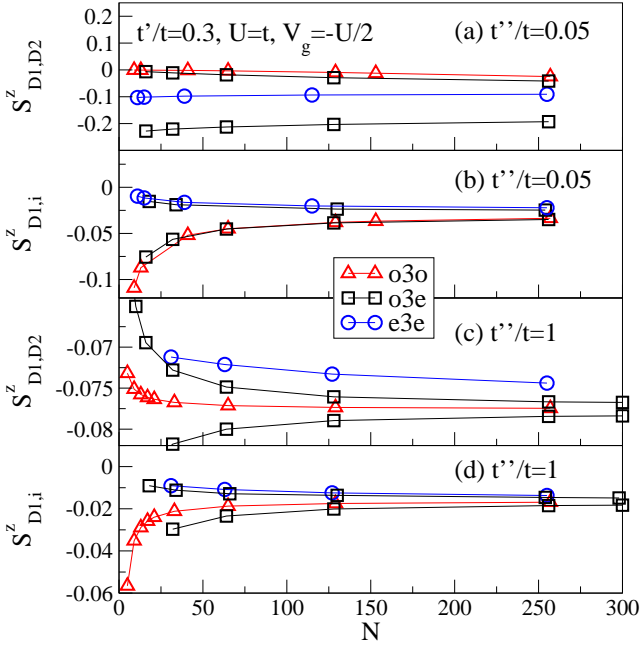}}
\caption{ Three dots, $t^{\prime}/t=0.3$, $U/t=1$.
Finite-size scaling of spin-spin correlations for $t^{\prime \prime}/t=0.05$ [(a),(b)]
and $t^{\prime \prime}/t=1$ [(c),(d)] (DMRG, $M\leq 500$). (a) and (c): spin-spin correlations between the central dot and the
outer ones. In the {\it o3e} case (squares), there are two branches as 
in reality, we alternate between {\it o3e} and {\it e3o} (even-dots-odd) configurations as $N$ increases. (b) and (d): spin-spin correlations between the outer dots and the
first site in the adjacent lead. In these two panels and for {\it o3e} clusters (squares), the upper(lower) curve
shows $S_{\mathbf{D1},i}$ ($S_{\mathbf{D3},i}$).}
\label{fig:3D_scaling_c}
\end{figure}

The rigid spins on either of these  cluster types imply suppressed spin fluctuations for
the individual dot ({\bf D2} in the case of {\it o3o} and {\bf D1}, {\bf D3} in the case of
{\it e3e})
as well as modified spin-spin correlations.
As two case examples, we present the finite-size scaling of such quantities with system size in
Figs.~\ref{fig:3D_scaling}  and \ref{fig:3D_scaling_c} for $U=t$, $t^{\prime}/t=0.3$ with  $t^{\prime \prime}/t=0.05$
and $t^{\prime \prime}/t=1$.

Let us first discuss {\it e3e} clusters
(circles in Figs.~\ref{fig:3D_scaling}  and \ref{fig:3D_scaling_c}).
Due to the large spin projection
on the outer two dots, spin fluctuations as well as the spin-spin correlations
with the first site in the leads are suppressed. The latter
is an indicator of how strongly the dots polarize the leads.
Since the spin-projections $S_{\mathrm{{\bf D1}}}^z$  and $S_{\mathrm{{\bf D3}}}^z$  are -- both
at small and large $t^{\prime \prime}$  -- only slowly
redistributed to other sites as the system size grows, as is evident from
Figs.~\ref{fig:3D_scaling}(a) and (c), we conclude that these clusters
will produce low conductance values or currents
 in  ECA and tDMRG calculations, respectively.

The {\it o3o} clusters (triangles in Figs.~\ref{fig:3D_scaling}  and \ref{fig:3D_scaling_c}) 
are the ones originally used in Ref.~\cite{busser04}. Figure~\ref{fig:3D_scaling}(b) reveals
that spin fluctuations on the central dot are suppressed in the
limit of small $t^{\prime \prime}\lesssim t$, which is due to a
large fraction of $S_{\mathrm{total}}^z$ located on the central dot.
This cluster type is therefore geometrically similar to the {\it
e1e} clusters of one quantum dot. As it is illustrated for  the case
of $t^{\prime \prime}/t=0.05$ in Fig.~\ref{fig:3D_scaling}(b), the
spin fluctuations on the central dot increase very slowly with
system size. We may therefore argue that here the central dot fails
to participate in a Kondo effect, especially in the two-stage Kondo
regime. This pathology, i.e., the finite
$S_{\mathrm{\bf D2}}^z$, is at the origin of the conductance dip at
$V_g=-U/2$ observed with ECA in Ref.~\cite{busser04}, similar
to the case of one quantum dot. 

Clusters with an overall $S_{\mathrm{total}}^z=0$ (i.e., {\it o3e} 
-- squares in Figs.~\ref{fig:3D_scaling}  and \ref{fig:3D_scaling_c})
exhibit the least significant finite-size effects and local
fluctuations converge the fastest with system size $N$, while
spin-spin correlations exhibit an asymmetry between the left and
right lead.
As a conclusion, we may expect   reliable results for
conductances from this cluster type, and this notion will further be
corroborated by ECA results in Sec.~\ref{sec:3D_eca}.


\subsection{ECA results for the conductance of three dots: Comparison of cluster types}
\label{sec:3D_eca}

\subsubsection{Gate-potential dependence}

\begin{figure}[t]
\centerline{\epsfig{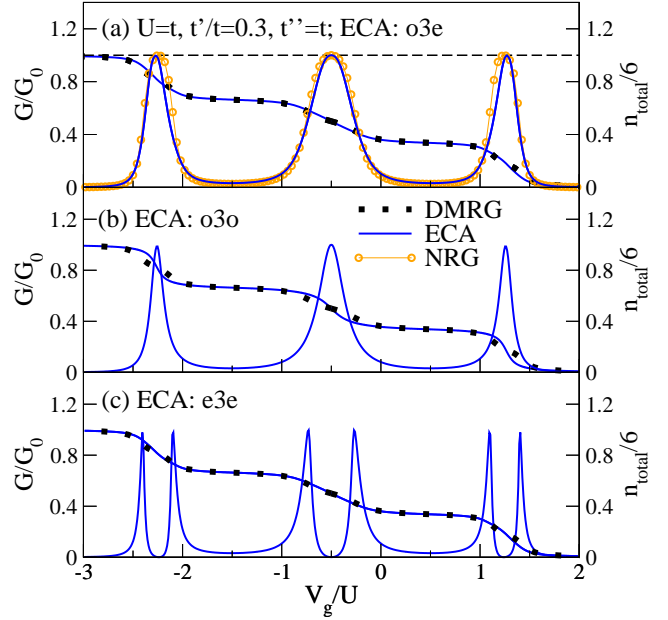}}
\caption{ Three dots, $t^{\prime}/t=0.3$, $U/t=1$, $t^{\prime \prime}/t=1$.
Conductance and charge vs. gate potential $V_g/U$.
(a) ECA results for the conductance and charge for an {\it o3e} 
cluster (solid lines, $N_{\mathrm{ED}}=12$)
vs. DMRG results for the  charge $n_{\mathrm{total}}$ (dotted line, $N=300$, $M=400$)
and  NRG results (\cite{zitko_unpub}, circles).
(b) ECA results for an {\it o3o} cluster (solid lines, $N_{\mathrm{ED}}=9$).
(c) ECA results for an {\it e3e}  cluster
(solid lines, $N_{\mathrm{ED}}=11$).
DMRG results for $n_{\mathrm{total}}$ (dotted line) are
included in (b) and (c) for comparison.
}\label{fig:3D_eca}
\end{figure}

In this section, we present ECA results for the conductance $G$ and $n_{\mathrm{total}}$ 
($n_{\mathrm{total}}=\langle n_{\mathrm{D1}}\rangle+\langle n_{\mathrm{D2}}\rangle+\langle n_{\mathrm{D3}}\rangle$) for the three
dots configuration as a function of the gate potential,
comparing all three cluster types. The ECA results (solid lines) are depicted in
Fig.~\ref{fig:3D_eca}(a) ({\it o3e}), (b) ({\it o3o}),  and (c) ({\it e3e}).
As expected from the discussion of
static properties in  Sec.~\ref{sec:3d_fs}, the results for $n_{\mathrm{total}}$ 
for {\it o3e} clusters obtained after embedding
agree very well with DMRG results [squares, see Fig.~\ref{fig:3D_eca}(a)], while
the results for the conductance from other clusters  deviate strongly
from DMRG data [squares, see Figs.~\ref{fig:3D_eca}(b) and (c)].
From Fig.~\ref{fig:3D_eca}(c) [solid lines], we realize that {\it e3e} clusters produce the expected
Coulomb-blockade like behavior due to the rigid spins on the outer dots (see Fig.~\ref{fig:3D_spin}).

 Figure~\ref{fig:3D_eca}(a)  also contains  NRG data from Ref.~\cite{zitko_unpub}. The agreement between ECA and 
NRG \cite{zitko_unpub} is quite good, especially for the central Kondo peak and
the Coulomb blockade valleys at $n_{\mathrm{total}}=2,4$. This clearly establishes that reliable results can be obtained from ECA calculations
using {\it o3e} clusters.

\subsubsection{Behavior at the particle-hole symmetric point}

Thus far, we have seen that with {\it o3e} clusters, we obtain the best agreement with NRG 
regarding the gate-potential dependence. Let us next address the
value of the conductance at the particle-hole symmetric point as a function
of $t^{\prime\prime}$.
This will indicate what  parameter range can be accessed
when decreasing $t^{\prime \prime}/t$ using {\it o3e} clusters.

\begin{figure}[t]
\centering
\centerline{\epsfig{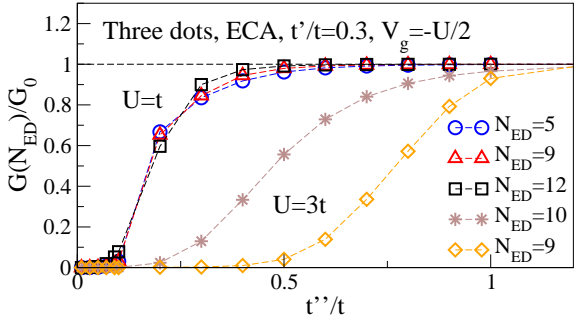}}
\caption{ 
Three dots with $t^{\prime}/t=0.3$, $U/t=1$, and $U/t=3$:
Variation of the conductance computed with ECA with $t^{\prime \prime}$ at the particle-hole symmetric point $V_g=-U/2$.
The results for $U/t=1.0$ are calculated using {\it o3o} with $N=5$ sites (circles) and $N=9$ sites (triangles), as well as
{\it o3e} clusters with $N=12$ sites (squares). At this value of $U/t$, there is barely any difference 
between the two types of clusters, i.e., as the value of $t^{\prime \prime}$ increases,
the conductance at the particle-hole symmetric point changes from 0 to $G_0$ at basically 
the same value of  $t^{\prime \prime}$, independent of cluster type.
However, the same is not true for the $U/t=3$ results ($N=10$ for {\it o3e} cluster (stars) and 
$N=9$ for {\it o3o} cluster (diamonds)). It is clear that the {\it o3o} clusters yield
a vanishing $G$  at substantially higher $t^{\prime \prime}$ than the {\it o3e} ones. 
As explained in the text, this is caused by a larger fixed spin in {\it o3o} clusters 
at higher $U$ values.
}
\label{fig2}
\end{figure}

Figure~\ref{fig2} compares {\it o3e} and {\it o3o} ECA conductance results 
at $V_g=-U/2$ for $U=t$ and $U=3t$. In the former case, 
where {\it o3o} clusters with $N_{\mathrm{ED}}=5,9$ are compared 
with an {\it o3e} cluster with $N_{\mathrm{ED}}=12$, it is shown that 
there is barely any difference between the results for the two different types of clusters: 
for $t^{\prime \prime}/t \approx 0.5$, the correct result $G/G_0=1$ is 
recovered. However, in the later case ($U=3t$), there is a pronounced 
difference between {\it o3o} ($N_{\mathrm{ED}}=9$) and {\it o3e} 
($N_{\mathrm{ED}}=10$) results: the vanishing of the conductance 
at the particle-hole symmetric point for the {\it o3o} cluster 
occurs at a much higher value of $t^{\prime \prime}/t$ than for 
the {\it o3e} cluster. The reason for that can be understood by 
analyzing the results in Fig.~\ref{fig:3D_spin}, where the fixed net-spin 
present in {\it o3o} clusters is compared for $U=t$ [panel (a)] 
and $U=3t$ [panel (b)]. A larger 
value of $U$, for a fixed $t^{\prime\prime}$, increases 
the magnitude of the fixed spins in the three dots (see
 Fig.~\ref{fig:3D_spin}). Therefore, as pointed out above, 
{\it o3e} clusters, with $S_{\mathrm{total}}^z=0$, provide 
better results at the particle-hole symmetric point 
when compared to {\it o3o} clusters. 
In the next subsection, we will discuss why the conductance eventually 
also vanishes for low $t^{\prime\prime}$ values in the 
{\it o3e} cluster and numerically accessible values of $L$, which do not exhibit finite  values of $\langle S_i^z\rangle $.

\begin{figure}[t]
\centerline{\epsfig{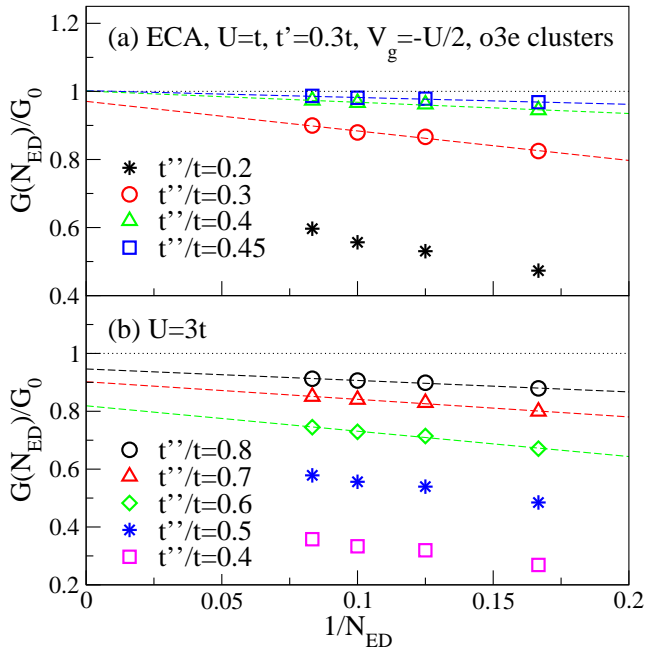}}
\caption{ Three dots, $t^{\prime}/t=0.3$, $U/t=1$ (a) and $U/t=3$ (b).
(a) Finite-size scaling analysis of ECA results on {\it o3e} clusters for $G(N_{\mathrm{ED}})$
at the particle-hole symmetric point ($V_g=-U/2$) for
$t^{\prime \prime}/t=0.45,0.4,0.3,0.2$ (squares, triangles, circles, and stars). 
Dashed lines are the linear fits to $G(N_{\mathrm{ED}})$.
(b) Same as in (a), but now for $U/t=3$, and $t^{\prime \prime}/t=0.8, 0.7, 0.6, 0.5,$ and 
$0.4$ (circles, triangles, diamonds, stars, and squares). Note that in this case, the 
threshold, below which the convergence to $G_0$ cannot be ascertained, is higher than 
in the $U/t=1$ case.
}\label{fig:3D_eca_fs}
\end{figure}

We now turn to a finite-size scaling analysis of $G(N)$ at $V_g=-U/2$.
Using cluster sizes within ECA that are numerically tractable (i.e.,
$N_{\mathrm{ED}} \lesssim 13$), we find that, at the particle-hole
symmetric point, for $U/t=1.0$, $G(N_{\mathrm{ED}})/G_0$ scales to one with
$1/N_{\mathrm{ED}}$ for $t^{\prime \prime}/t\gtrsim 0.3$ (upper panel in Fig.~\ref{fig:3D_eca_fs}).
In addition, it is clear from the results shown
in Fig.~\ref{fig:3D_eca_fs}~(b) that, the larger the
value of $U$, the higher is the value of $t^{\prime \prime}$ below
which one cannot clearly ascertain the extrapolation of $G(N_{\mathrm{ED}})$ to
$G_0$ at small $t^{\prime \prime}$ values.
Obviously, this happens because the Kondo temperature of the second stage decreases with $U$,  causing the ECA conductance to 
vanish at a larger $t^{\prime\prime}$. 
Summarizing, both {\it o3o} and {\it o3e} clusters
produce results in qualitative agreement with other techniques
at $V_g=-U/2$ above certain values of $t^{\prime \prime}$.
For $U/t=1$, $G=G_0$ can be obtained by extrapolation with a $G(N_{\mathrm{ED}})=G_0-\mathrm{const}/N_{\mathrm{ED}}$ function for $t^{\prime\prime}/t\gtrsim 0.2$.
For smaller $t^{\prime \prime}$, system sizes are  too small to
establish a clear trend of $G(N_{\mathrm{ED}})/G_0$ towards  unity due to a
finite level spacing. 

We are led to conclude that ECA captures quite well the
correct physics depending upon the parameters and the cluster taken.
The optimum cluster can be identified by analyzing the finite-size scaling of, e.g., 
the charge, with either the Lanczos solver that is part of ECA or independent techniques such as ground state DMRG, the latter being available in
open access releases, of, e.g., the ALPS project \cite{albuquerque07}.

\subsection{{\it o3e} clusters in the small $t^{\prime\prime}$ regime}
\label{sec:o3e}
We next wish to illustrate the finite-size effects encountered in the small $t^{\prime\prime}$ regime, for {\it o3e} clusters.
Unfortunately,
when $t^{\prime\prime}$ is decreased, the asymmetry between the two outer dots on
{\it o3e} becomes quite pronounced. Note that  the total charge from
ECA still agrees well with DMRG down to very small $t^{\prime\prime}$ (not shown). However,
spin-spin correlations on small {\it o3e} clusters suffer from
{\it (i)} the aforementioned asymmetry between the two outer dots (see Fig.~\ref{fig:3D_scaling_c}), and
{\it (ii)} the fact that the central dot is basically decoupled from the leads 
(at the Fermi energy). This is  indicated by small spin-spin
correlations between the central dot and the first site in the
leads, which indicates that the effective exchange with the leads is 
smaller than the energy level spacing.
Consistent with the first point discussed here, in accordance to the LDOS calculated with ECA for 
an {\it o3e} configuration (not shown), at small $t^{\prime\prime}$ and for $V_g=-U/2$ (particle-hole symmetric point), 
only one of the outer dots develops a sharp Kondo resonance, namely the one 
connected to the {\it odd} lead. The dot connected to the {\it even} lead 
can only be Kondo screened through the other two dots. However, at low 
values of $t^{\prime \prime}$, this mechanism becomes ineffective and 
thus the dot's LDOS at the Fermi energy is low, 
leading to the suppression of the conductance at the particle-hole 
symmetric point, as seen in Fig.~\ref{fig2}. 
Therefore, even for the optimum cluster type {\it o3e}, the two-stage Kondo regime  seems to be out of 
reach within the real-space variant of ECA and current computational resources. 

\section{Summary}
\label{conc}

In this work, we have performed an extensive finite-size scaling
analysis of fluctuations and correlations in nanostructures, such as
one and three quantum dots coupled to noninteracting, tight-binding leads.
Strong differences in the finite-size scaling behavior emerge
depending on even-odd effects \cite{sorensen96}. These findings are relevant for the
interpretation of results from numerical approaches for the calculation of the
conductance of nanostructures that are based on analyzing  clusters
of finite length with open boundary conditions, such as
time-dependent DMRG or the embedded-cluster approximation (ECA). To
qualitatively capture precursors of the Kondo effect within these two methods,   clusters with a
$S_{\mathrm{total}}^z=0$ are best suited,  as electrons on the dot
at half filling naturally participate in a singlet state. Other
cluster types with an overall $S_{\mathrm{total}}^z=1/2$ may be {\it
far} away from capturing Kondo physics in the sense of  slow
convergence with system size. For certain configurations,
 most of the $S_{\mathrm{total}}^z=1/2$ is
found on the central site, which, if this site is a quantum dot,
will cause a gap in the local density of states at the Fermi level,
reminiscent of the Zeeman-splitting of the conductance due to the
application of a magnetic field.

 These findings  help to understand the  discrepancy between ECA results for three
dots and those of other techniques: the  vanishing of the conductance at the particle-hole symmetric point
obtained with ECA is only seen on certain cluster types, but disappears when the three dots
 are embedded in a cluster with an overall $S_{\mathrm{total}}^z=0$. 
 In the latter case, the ECA results are then in agreement with the picture proposed in 
 Refs.~\cite{karrasch06,zitko06,oguri99,oguri05,zitko07,nisikawa06,lobos06}.
 While
 parameter  regimes  with exponentially small energy scales such as two-stage Kondo
 regimes are difficult to be accessed with this
 technique in its real-space variant, very good agreement with exact results such as the Friedel sum rule
 in the case of one dot -- as shown in Fig.~\ref{fig:1dot_eca}(a) --
 or NRG in the case of three dots is otherwise found in other regions of the parameter space. The latter is
 established from Fig.~\ref{fig:3D_eca}(a) and Fig.~\ref{fig2}.

Another model that exhibits a two-stage Kondo effect is   two dots coupled in a T-shape geometry \cite{cornaglia05,zitko06b}
At half filling, the conductance is found to vanish \cite{cornaglia05,zitko06b}, a picture that disagrees  with ECA results in the two-stage Kondo regime \cite{busser04b}. 
We have performed a similar analysis of static properties and ECA results on different cluster types for this geometry, with results not shown here. While in the regime of large $t^{\prime\prime}/U$,
 where $t^{\prime\prime}$ is the hopping between the two dots, good agreement between the Friedel sum rule \cite{cornaglia05}
 and ECA is found, the two-stage Kondo regime, which emerges at small $t^{\prime\prime}$ \cite{cornaglia05,zitko06b}
 suffers from strong finite-size effects. Unfortunately, no cluster type shows a  particularly fast
 scaling behavior here. 

We conclude that the determination of the optimum cluster type is crucial. This can be 
 done by studying different cluster types with the Lanczos solver that is part of standard
 ECA codes.
 Further, a  comparison of ECA results for static
 properties, such as the total charge of a nanostructure as a function of gate potential and very importantly, spin and charge
 fluctuations,
 against independent techniques, such as ground-state  DMRG, is helpful in this respect as well, as we showed in this work.
 Promising results from a new version of ECA that incorporates a logarithmic discretization 
of the DOS of the leads clearly indicate that the range of validity
 of the embedded-cluster approximation can further be  improved \cite{anda08}.


 \begin{acknowledgement}
 It is a pleasure to thank   Hsiu-Hau Lin, Volker Meden, 
Jose Riera, and Marcos Rigol for fruitful
 discussions. We thank Rok {\v{Z}}itko for providing us with NRG data for three dots. 
 E.V.A. acknowledges support from   CNPq and FAPERJ, Brazil. G.B.M. acknowledges support
from NSF (DMR-0710529) and Research Corporation (Contract No. CC6542). 
 Research at ORNL is sponsored by the Division of Materials Sciences and Engineering,
Office of Basic Energy Sciences, U.S. Department of Energy, under contract
DE-AC05-00OR22725 with Oak
Ridge National Laboratory, managed and operated by UT-Battelle, LLC.
K.A.A., E.D., and F.H.-M.  are supported in part by  NSF grant DMR-0706020. C.A.B acknowledges support from NSF grant DMR-0710529.
\end{acknowledgement}


\bibliographystyle{/net/home/lxtsfs1/tpc/fabi/bibliothek/pf}
\bibliography{/net/home/lxtsfs1/tpc/fabi/bibliothek/bbase}

\begin{thebibliography}{10}

\bibitem{goldhabergordon98}
D.~Goldhaber-Gordon, H.~Shtrikman, D.~Mahalu, D.~Abusch-Magder, U.~Meirav, and
  M.~A. Kastner,
\newblock Nature {\bf 391}, 156 (1998).

\bibitem{glazman88}
L.~Glazman and M.~Raikh,
\newblock JETP Lett. {\bf 47}, 452 (1988).

\bibitem{ng88}
T.~K. Ng and P.~A. Lee,
\newblock Phys. Rev. Lett. {\bf 61}, 1768 (1988).

\bibitem{meir92}
Y.~Meir and P.~Lee,
\newblock Phys. Rev. Lett. {\bf 68}, 2512 (1992).

\bibitem{glazman05}
L.~Glazman and M.~Pustilnik,
\newblock {\em Nanophysics: Coherence and Transport, eds. H. Bouchiat et al.},
  p. 427,
\newblock Elsevier, 2005.

\bibitem{pustilnik05}
M.~Pustilnik,
\newblock phys. stat. sol. (a) {\bf 203}, 1137 (2006).

\bibitem{grobis07}
M.~Grobis, I.~Rau, R.~Potok, and D.~Goldhaber-Gordon,
\newblock {\em Kondo Effect in Mesoscopic Quantum Dot},
\newblock Wiley, 2007.

\bibitem{wiel00}
W.~G. van~der Wiel, S.~D. Franceschi, T.~Fujisawa, J.~M. Elzerman, S.~Tarucha,
  and L.~P. Kouwenhoven,
\newblock Science {\bf 289}, 2105 (2000).

\bibitem{park02}
J.~Park, A.~N. Pasupathy, J.~I. Goldsmith, C.~Chang, Y.~Yaish, J.~R. Petta,
  M.~Rinkoski, J.~P. Sethna, H.~D. Abruna, P.~L. McEuen, and D.~C. Ralph,
\newblock Nature {\bf 417}, 722 (2002).

\bibitem{wilson75}
K.~G. Wilson,
\newblock Rev. Mod. Phys. {\bf 47}, 773 (1975).

\bibitem{krishnamurthy80a}
H.~R. Krishna-murthy, J.~W. Wilkins, and K.~G. Wilson,
\newblock Phys. Rev. B {\bf 21}, 1003 (1980).

\bibitem{anders05}
F.~B. Anders and A.~Schiller,
\newblock Phys.\ Rev.\ Lett. {\bf 95}, 196801 (2005).

\bibitem{bulla08}
R.~Bulla, T.~Costi, and T.~Pruschke,
\newblock Rev. Mod. Phys. {\bf 80}, 395 (2008).

\bibitem{bohr06}
D.~Bohr, P.~Schmitteckert, and P.~W\"olfle,
\newblock Europhys. Lett. {\bf 73}, 246 (2006).

\bibitem{bohr07}
D.~Bohr and P.~Schmitteckert,
\newblock Phys. Rev. B {\bf 75}, 241103(R) (2007).

\bibitem{weichselbaum08}
A.~Weichselbaum, F.~Verstraete, U.~Schollw\"ock, J.~I. Cirac, and J.~von Delft,
\newblock cond-mat/0504305  (unpublished).

\bibitem{alhassanieh06}
K.~A. Al-Hassanieh, A.~E. Feiguin, J.~A. Riera, C.~A. B\"usser, and E.~Dagotto,
\newblock Phys. Rev. B {\bf 73}, 195304 (2006).

\bibitem{schneider06}
G.~Schneider and P.~Schmitteckert,
\newblock cond-mat/0601389  (unpublished).

\bibitem{kirino08}
S.~Kirino, T.~Fujii, J.~Zhao, and K.~Ueda,
\newblock J. Phys. Soc. Jpn. {\bf 77}, 084704 (2008).

\bibitem{feiguin08c}
A.~Feiguin, P.~Fendley, M.~P. Fisher, and C.~Nayak,
\newblock arXiv:0809.1415 (unpublished).

\bibitem{boulat08}
E.~Boulat, H.~Saleur, and P.~Schmitteckert,
\newblock Phys. Rev. Lett. {\bf 101}, 140601 (2008).

\bibitem{dasilva08}
L.~G. G. V.~D. da~Silva, F.~Heidrich-Meisner, A.~E. Feiguin, C.~A. B\"usser,
  G.~B. Martins, E.~V. Anda, and E.~Dagotto,
\newblock Phys. Rev. B, {in press}, arXiv:0807.0581.

\bibitem{weiss08}
S.~Weiss, J.~Eckel, M.~Thorwart, and R.~Egger,
\newblock Phys. Rev. B {\bf 77}, 195316 (2008).

\bibitem{kehrein05}
S.~Kehrein,
\newblock Phys.\ Rev.\ Lett. {\bf 95}, 056602 (2005).

\bibitem{ferrari99}
V.~Ferrari, G.~Chiappe, E.~V. Anda, and M.~A. Davidovich,
\newblock Phys. Rev. Lett. {\bf 82}, 5088 (1999).

\bibitem{busser00}
C.~A. B\"usser, E.~V. Anda, A.~L. Lima, M.~A. Davidovich, and G.~Chiappe,
\newblock Phys. Rev. B {\bf 62}, 9907 (2000).

\bibitem{chiappe03}
G.~Chiappe and J.~A. Verges,
\newblock J. Phys.: Condens. Matter {\bf 15}, 8805 (2003).

\bibitem{busser04}
C.~A. B\"usser, A.~Moreo, and E.~Dagotto,
\newblock Phys. Rev. B {\bf 70}, 035402 (2004).

\bibitem{molina03}
R.~Molina, D.~Weinmann, R.~Jalabert, G.-L. Ingold, and J.-L. Pichard,
\newblock Phys. Rev. B {\bf 67}, 235306 (2003).

\bibitem{rejec03}
T.~Rejec and A.~Ram\ifmmode~\check{s}\else \v{s}\fi{}ak,
\newblock Phys. Rev. B {\bf 68}, 035342 (2003).

\bibitem{meden03}
V.~Meden and U.~Schollw\"ock,
\newblock Phys. Rev. B {\bf 67}, 193303 (2003).

\bibitem{kaminski00}
A.~Kaminski, Y.~V. Nazarov, and L.~I. Glazman,
\newblock Phys. Rev. B {\bf 62}, 8154 (2000).

\bibitem{rosch01}
A.~Rosch, J.~Kroha, and P.~W\"olfle,
\newblock Phys. Rev. Lett. {\bf 87}, 156802 (2001).

\bibitem{doyon06}
B.~Doyon and N.~Andrei,
\newblock Phys. Rev. B {\bf 73}, 245326 (2006).

\bibitem{schoeller00}
H.~Schoeller and J.~K\"onig,
\newblock Phys.\ Rev.\ Lett. {\bf 84}, 3686 (2000).

\bibitem{karrasch06}
C.~Karrasch, T.~Enss, and V.~Meden,
\newblock Phys. Rev. B {\bf 73}, 235337 (2006).

\bibitem{jakobs07}
S.~G. Jakobs, V.~Meden, and H.~Schoeller,
\newblock Phys. Rev. Lett. {\bf 99}, 150603 (2007).

\bibitem{kotliar86}
G.~Kotliar and A.~E. Ruckenstein,
\newblock Phys. Rev. Lett. {\bf 57}, 1362 (1986).

\bibitem{schoenhammer85}
O.~Gunnarsson and K.~Sch\"onhammer,
\newblock Phys. Rev. B {\bf 31}, 4815 (1985).

\bibitem{zitko06}
R.~{\v{Z}}itko, J.~Bon{\v{c}}a, A.~Ram{\v{s}}ak, and T.~Rejec,
\newblock Phys. Rev. B {\bf 73}, 153307 (2006).

\bibitem{marques06}
M.~A. L.~Marques et~al. (Eds.),
\newblock Lect. Notes. Phys. {\bf 706} (2006).

\bibitem{koentopp06}
M.~Koentopp, K.~Burke, and F.~Evers,
\newblock Phys. Rev. B {\bf 73}, 121403 (2006).

\bibitem{arnold07}
A.~Arnold, F.~Weigend, and F.~Evers,
\newblock J. Chem. Phys. {\bf 126}, 174101 (2007).

\bibitem{schmitteckert08}
P.~Schmitteckert and F.~Evers,
\newblock Phys. Rev. Lett. {\bf 100}, 086401 (2008).

\bibitem{brandbyge02}
M.~Brandbyge, J.-L. Mozos, P.~Ordej\'on, J.~Taylor, and K.~Stokbro,
\newblock Phys. Rev. B {\bf 65}, 165401 (2002).

\bibitem{alhassanieh05}
K.~A. Al-Hassanieh, C.~A. B\"usser, G.~B. Martins, and E.~Dagotto,
\newblock Phys. Rev. Lett. {\bf 95}, 256807 (2005).

\bibitem{martins06}
G.~B. Martins, C.~A. B\"usser, K.~A. Al-Hassanieh, E.~V. Anda, A.~Moreo, and
  E.~Dagotto,
\newblock Phys.\ Rev.\ Lett. {\bf 76}, 066802 (2006).

\bibitem{busser07}
C.~A. B\"{u}sser and G.~B. Martins,
\newblock Phys. Rev. B {\bf 75}, 045406 (2007).

\bibitem{martins05}
G.~B. Martins, C.~A. B\"{u}sser, K.~A. Al-Hassanieh, A.~Moreo, and E.~Dagotto,
\newblock Phys. Rev. Lett. {\bf 94}, 026804 (2005).

\bibitem{white92b}
S.~R. White,
\newblock Phys.\ Rev.\ Lett. {\bf 69}, 2863 (1992).

\bibitem{white93}
S.~R. White,
\newblock Phys.\ Rev.\ B {\bf 48}, 10345 (1993).

\bibitem{schollwoeck05}
U.~Schollw\"ock,
\newblock Rev. Mod. Phys. {\bf 77}, 259 (2005).

\bibitem{hallberg06}
K.~Hallberg,
\newblock Advances in Physics {\bf 55}, 477 (2006).

\bibitem{gubernatis87}
J.~E. Gubernatis, J.~E. Hirsch, and D.~J. Scalapino,
\newblock Phys. Rev. B {\bf 35}, 8478 (1987).

\bibitem{thimm99}
W.~B. Thimm, J.~Kroha, and J.~von Delft,
\newblock Phys. Rev. Lett. {\bf 82}, 2143 (1999).

\bibitem{sorensen05}
E.~S. S{\o}rensen and I.~Affleck,
\newblock Phys.\ Rev.\ Lett. {\bf 94}, 086601 (2005).

\bibitem{costamagna06}
S.~Costamagna, C.~Gazza, M.~Torio, and J.~A. Riera,
\newblock Phys. Rev. B {\bf 74}, 195103 (2006).

\bibitem{borda06}
L.~Borda,
\newblock Phys. Rev. B {\bf 75}, 041307 (2007).

\bibitem{andrei80}
N.~Andrei,
\newblock Phys. Rev. Lett. {\bf 45}, 379 (1980).

\bibitem{gerland00}
U.~Gerland, J.~von Delft, T.~A. Costi, and Y.~Oreg,
\newblock Phys. Rev. Lett. {\bf 84}, 3710 (2000).

\bibitem{hewson}
A.~Hewson,
\newblock {\em The Kondo Problem to Heavy Fermions},
\newblock Cambridge, UK, 1993.

\bibitem{sorensen96}
E.~S. S{\o}rensen and I.~Affleck,
\newblock Phys. Rev. B {\bf 53}, 9153 (1996).

\bibitem{kim02}
T.-S. Kim and S.~Hershfield,
\newblock Phys. Rev. B {\bf 65}, 214526 (2002).

\bibitem{hand06}
T.~Hand, J.~Kroha, and H.~Monien,
\newblock Phys. Rev. Lett. {\bf 97}, 136604 (2006).

\bibitem{oguri99}
A.~Oguri,
\newblock Phys. Rev. B {\bf 59}, 12240 (1999).

\bibitem{oguri05}
A.~Oguri and A.~C. Hewson,
\newblock J. Phys. Soc. Jpn. {\bf 74}, 988 (2005).

\bibitem{zitko07}
R.~{\v{Z}}itko and J.~Bon{\v{c}}a,
\newblock Phys. Rev. Lett. {\bf 98}, 047203 (2007).

\bibitem{nisikawa06}
Y.~Nisikawa and A.~Oguri,
\newblock Phys. Rev. B {\bf 73}, 125108 (2006).

\bibitem{lobos06}
A.~M. Lobos and A.~A. Aligia,
\newblock Phys. Rev. B {\bf 74}, 165417 (2006).

\bibitem{kuzmenko03}
T.~Kuzmenko, K.~Kikoin, and Y.~Avishai,
\newblock Europhys. Lett. {\bf 64}, 218 (2003).

\bibitem{white04}
S.~R. White and A.~E. Feiguin,
\newblock Phys.\ Rev.\ Lett. {\bf 93}, 076401 (2004).

\bibitem{daley04}
A.~Daley, C.~Kollath, U.~Schollw\"ock, and G.~Vidal,
\newblock J. Stat. Mech.: Theory Exp. , P04005 (2004).

\bibitem{anda81}
E.~V. Anda,
\newblock J. Phys. C {\bf 14}, 1037 (1981).

\bibitem{metzner91}
W.~Metzner,
\newblock Phys. Rev. B {\bf 43}, 8549 (1991).

\bibitem{davidovich02}
M.~A. Davidovich, E.~V. Anda, C.~A. B\"usser, and G.~Chiappe,
\newblock Phys. Rev. B {\bf 65}, 233310 (2002).

\bibitem{anda02}
E.~V. Anda, C.~A. B\"usser, G.~Chiappe, and M.~A. Davidovich,
\newblock Phys. Rev. B {\bf 66}, 035307 (2002).

\bibitem{chiappe05}
G.~Chiappe, E.~Louis, E.~V. Anda, and J.~A. Verges,
\newblock Phys. Rev. B {\bf 71}, 241405(R) (2005).

\bibitem{aguiar-hualde07}
J.~M. Aguiar-Hualde, G.~Chiappe, E.~Louis, and E.~V. Anda,
\newblock Phys. Rev. B {\bf 76}, 155427 (2007).

\bibitem{busser04b}
C.~A. B\"usser, G.~B. Martins, K.~Al-Hassanieh, A.~Moreo, and E.~Dagotto,
\newblock Phys. Rev. B {\bf 70}, 245303 (2004).

\bibitem{wingreen93}
N.~S. Wingreen, A.-P. Jauho, and Y.~Meir,
\newblock Phys. Rev. B {\bf 48}, 8487 (1993).

\bibitem{schiller00}
A.~Schiller and S.~Hershfield,
\newblock Phys. Rev. B {\bf 62}, R16271 (2000).

\bibitem{cazalilla02}
M.~A. Cazalilla and J.~B. Marston,
\newblock Phys. Rev. Lett. {\bf 88}, 256403 (2002).

\bibitem{alhassanieh-unpub}
K.~A. Al-Hassanieh et~al.,
\newblock   unpublished results.

\bibitem{anda08}
E.~V. Anda, G.~Chiappe, C.~A. B\"usser, M.~A. Davidovich, G.~B. Martins,
  F.~Heidrich-Meisner, and E.~Dagotto,
\newblock Phys. Rev. B {\bf 78}, 085308 (2008).

\bibitem{lieb68}
E.~Lieb and F.~Y. Wu,
\newblock Phys. Rev. Lett. {\bf 20}, 1445 (1968).

\bibitem{cornaglia05}
P.~S. Cornaglia and D.~R. Grempel,
\newblock Phys. Rev. B {\bf 71}, 075305 (2005).

\bibitem{zitko_unpub}
R.~{\v{Z}}itko,
\newblock unpublished results.

\bibitem{albuquerque07}
A.~Albuquerque, F.~Alet, P.~Corboz, P.~Dayal, A.~Feiguin, S.~Fuchs, L.~Gamper,
  E.~Gull, S.~Guertler, A.~H. an~R.~Igarashi, M.~Koerner, A.~Kozhevnikov,
  A.~Laeuchli, S.~Manmana, M.~Matsumoto, I.~McCulloch, F.~Michel, R.~Noack,
  G.~Pawlowski, L.~Pollet, T.~Pruschke, U.~Schollw\"ock, S.~Todo, S.~Trebst,
  M.~Troyer, P.~Werner, and S.~Wessel,
\newblock J. Mag. Mag. Mat. {\bf 310}, 1187 (2007).

\bibitem{zitko06b}
R.~{\v{Z}}itko and J.~Bon{\v{c}}a,
\newblock Phys. Rev. B {\bf 73}, 035332 (2006).

\end{thebibliography}

\end{document}